\documentclass[
aps,
prb,
showpacs,
floatfix,
reprint,
twoside,
superscriptaddress
]{revtex4-1}
\usepackage{amsmath}
\usepackage[T1]{fontenc}
\usepackage{fourier}
\usepackage{graphicx}
\usepackage[colorlinks=true,
allcolors=blue,
dvipdfm=true]{hyperref}

\begin{document}

\title{ac properties of short Josephson weak links}

\author{Andreas Moor}
\affiliation{Theoretische Physik III, Ruhr-Universit\"{a}t Bochum, D-44780 Bochum, Germany}
\author{Anatoly F.~Volkov}
\affiliation{Theoretische Physik III, Ruhr-Universit\"{a}t Bochum, D-44780 Bochum, Germany}

\begin{abstract}
We calculate the admittance of two types of Josephson weak links---the first
is a one-dimensional superconducting wire with a local suppression of the
order parameter, and the second is a short S-c-S structure, where~S denotes
a superconducting reservoir and~c is a constriction. The systems of the
first type are analyzed on the basis of time-dependent Ginzburg-Landau
equations derived by Gor'kov and Eliashberg for gapless superconductors with
paramagnetic impurities. It is shown that the impedance~$Z(\Omega)$ has a
maximum as a function of the frequency~$\Omega$, and the electric field~$%
E_{\Omega}$ is determined by two gauge-invariant quantities. One of them is
the condensate momentum~$Q_{\Omega}$ and another is a potential~$\mu$
related to charge imbalance. The structures of the second type are studied
on the basis of microscopic equations for quasiclassical Green's functions
in the Keldysh technique. For short S-c-S contacts (the Thouless energy~${E_{%
\text{Th}} = D/L^{2} \gg \Delta}$) we present a formula for admittance~$Y$
valid frequencies~$\Omega$ and temperatures~$T$ less than the Thouless
energy~$E_{\text{Th}}$ (${\hbar \Omega, T \ll E_{\text{Th}}}$) but arbitrary
with respect to the energy gap~$\Delta$. It is shown that, at low
temperatures, the absorption is absent [${\mathrm{Re}(Y) = 0}$] if the
frequency does not exceed the energy gap in the center of the constriction ($%
{\Omega < \Delta \cos \varphi_{0}}$, where~$2 \varphi_{0}$ is the phase
difference between the S~reservoirs). The absorption gradually increases
with increasing the difference ${(\Omega - \Delta \cos \varphi_{0})}$ if~$2
\varphi_{0}$ is less than the phase difference~$2 \varphi_{\text{c}}$
corresponding to the critical Josephson current. In the interval ${2
\varphi_{\text{c}} < 2 \varphi_{0} < \pi}$, the absorption has a maximum.
This interval of the phase difference is achievable in phase-biased
Josephson junctions. Close to~$T_{\text{c}}$ the admittance has a maximum at
low~$\Omega$ which is described by an analytical formula.
\end{abstract}

\date{\today}
\pacs{}
\maketitle

\affiliation{Theoretische Physik III, Ruhr-Universit\"{a}t Bochum, D-44780
Bochum, Germany} 

\affiliation{Theoretische Physik III, Ruhr-Universit\"{a}t Bochum, D-44780
Bochum, Germany}

\section{Introduction}

The study of dynamic effects in superconductors began soon after the
appearance of microscopic BCS theory of superconductivity.\cite{BCS} Using
the BCS theory, Mattis and Bardeen have calculated the admittance of a
superconductor~$Y(\Omega, q)$.\cite{Mattis_Bardeen_1958} Later, Abrikosov,
Gor'kov and Khalatnikov have obtained the admittance for pure
superconductors by using the Green's function technique.\cite%
{Abrikosov_Gorkov_Khalatnikov_1958} This technique was applied by Abrikosov
and Gor'kov to calculate the linear response of superconductors with
impurities.\cite{AbrGor58} In more detail, the theory of admittance has been
later developed by Nam.\cite{Nam67,*Nam67_a} In these papers, it has been
shown that at low temperatures absorption is absent if the frequency of
electromagnetic field~$\Omega$ is less than~$2 \Delta$. This means that the
real part of admittance~${\mathrm{Re}[Y(\Omega)] \equiv Y^{\prime}(\Omega)}$
equals zero in the limit ${T \to 0}$ and ${\Omega < 2 \Delta/\hbar}$. If
frequency~$\Omega$ exceeds~$2 \Delta$, $Y^{\prime}(\Omega)$~increases with
increasing the difference ${(\Omega - 2 \Delta)}$.

On the other hand, the intensive study of dynamic collective modes in
superconductors, both in low- and high\nobreakdash-T$_{\text{c}}$ ones, is
carried out in the last decade. A special attention is paid to the amplitude
mode (AM), which is called often in literature the Higgs mode.\cite{Higgs64}
This mode has been studied theoretically long ago\cite%
{VolkovKogan73,Kulik_et_al_1981,Levitov04,*Levitov04a,Amin_et_al_2004,Warner_Leggett_2005,Simons05,Altshuler05,*Altshuler05a,Yuzbashyan06,Yuzbashyan06a,Gurarie09,Manske14,Moor14,Levchenko15,Yuzbashyan_et_al_2015,Peronaci15,Kemper_et_al_2015,Aoki15,Cea_et_al_2015,Cea_et_al_2016,Murakami_et_al_2016_03,Kemper16,Chou_Liao_Foster_2016_arxiv,Yoon_Watanabe_2015_arxiv}%
, but only recently it was observed in experiments.\cite%
{ExpS13,Matsunaga_et_al_2014} A superconductor (Nb$_{1-x}$Ti$_{x}$N) was
driven out of the equilibrium by a short laser pulse (teraherz frequency
range) and the temporal evolution of the deviation~$\delta \Delta(t)$ from
the equilibrium value~$\Delta$ was detected by a weak probe signal in
picosecond time interval. This evolution can be qualitatively described by
the equation\cite{VolkovKogan73}
\begin{equation}
\delta \Delta(t) \propto \delta \Delta(0) \frac{\cos (2 \Delta t / \hbar )}{%
\sqrt{2 \Delta t / \hbar}} \,.  \label{I-1}
\end{equation}

A weak incident electric field ${\mathbf{E}(t) = \mathbf{E}_{\Omega}
\cos(\Omega t)}$ obviously can not lead to a perturbation of the order
parameter~$\Delta$ because it is a scalar so that~$\delta \Delta(t)$ can be
proportional only to even orders of $\mathbf{E}^{2n}(t)$. However, as we
have shown recently,~\cite{Moor_Volkov_Efetov_2017_PRL} the situation
changes in the presence of the condensate flow. In this case, even a weak ac
field~$\mathbf{E}(t)$ leads to a perturbation of~$\Delta$, ${\delta
\Delta_{\Omega} \propto \mathbf{Q}_{\Omega} \mathbf{Q}_{0}}$, where ${%
\mathbf{Q}_{0} = m \mathbf{v}_{0}}$ is the condensate momentum,~$\mathbf{v}%
_{0}$ is the velocity of the condensate, and~$\mathbf{Q}_{\Omega}$ is the ac
condensate momentum induced by the electric field~$\mathbf{E}_{\Omega}$
according to the expression
\begin{equation}
-i \Omega \hbar \mathbf{Q}_{\Omega} = e \mathbf{E}_{\Omega} \,.  \label{I0}
\end{equation}

If the frequency of the external electric field~$\Omega$ coincides with the
frequency of the AM~$2 \Delta /\hbar$, a resonance absorption of the
incident electromagnetic field~$\mathbf{E}_{\Omega}$ takes place and the
real part~${Y^{\prime}(\Omega) \equiv \mathrm{Re}[Y(\Omega)]}$ of admittance
has a sharp peak at ${\Omega = 2 \Delta / \hbar}$.

A similar peak was obtained in Ref.~\onlinecite{Dai_Lee_2017}, where linear
response of a superconductor with a finite-momentum pairing was calculated.
As the authors of Ref.~\onlinecite{Dai_Lee_2017} claim, their results can be
applied to high\nobreakdash-$T_{\text{c}}$ superconductors with a pair
density wave or to superconductors in the
Fulde-Ferrell-Larkin-Ovchinnikov~(FFLO) state.~\cite%
{Fulde_Ferrell_1964,Larkin_Ovchinnikov_1965} In both cases, the
superconducting order parameter depends on coordinate,~$\Delta (\mathbf{r})$%
, turning to zero at some points or lines.

High frequency properties of superconductors are important not only from the
point of view of fundamental physics, but also of applications. In
particular, the use of superconducting devices in qubits and in highly
sensitive detectors requires the knowledge of the admittance~$Y(\Omega)$.%
\cite%
{Goltsman_et_al_2001,Day_et_al_2003,Janssen_et_al_2013,Clarke_Wilhelm_2008}
The systems used in practical devices often include Josephson
junctions~(JJ), for example, S\nobreakdash-c\nobreakdash-S or S\nobreakdash-n%
\nobreakdash-S weak links of different types, where~c denotes a constriction
and~n stands for a normal metal. The study of ac properties of JJs has began
long ago (see references in Refs.~\onlinecite{Likharev,Barone_Paterno_1982}%
). The admittance~$Y(\Omega)$ of a short JJ of the S\nobreakdash-c%
\nobreakdash-S type has been calculated by Artemenko~\emph{et al.} on the
basis of Keldysh technique for quasiclassical Green's functions.\cite{AVZ79}
It was assumed that the Thouless energy~${E_{\text{Th}} = D / L^{2}}$ is
much larger than~$T_{\text{c}}$. In particular, it was shown that at low
frequencies~$\Omega$ and close to the critical temperature~$T_{\text{c}}$
the admittance has the form [see Eq.~(31) in Ref.~\onlinecite{AVZ79}]
\begin{equation}
Y(\Omega) = \frac{2 e I_{\text{c}}(\nu_{\text{in}} + i \Omega)}{\hbar
(\Omega^{2} + \nu_{\text{in}}^{2})} P(2 \varphi_{0}) + \frac{1}{R} \,,
\label{I1}
\end{equation}
where ${I_{\text{c}} = \pi \Delta^{2} / (4 e T R)}$ is the critical current
of this JJ near~T$_{\text{c}}$, $R$~is the resistance in the normal state
and~$\nu_{\text{in}}$ is inelastic scattering time.\cite%
{Kulik_Omelyanchuk_1978} The function~$P(2 \varphi_{0})$ is a function of
the phase difference~$2 \varphi_{0}$. The form of~$P(\varphi_{0})$ is
displayed in Fig.~\ref{fig:P}. Equation~(\ref{I1}) shows that the reactive
part of admittance has a sharp peak at a small frequency ${\Omega \simeq
\nu_{\text{in}}}$ since ${\nu_{\text{in}} \ll \Delta}$.

An anomalous behavior of the admittance~$Y(\Omega)$ was obtained also in
Ref.~\onlinecite{Kos_Nigg_Glazman_2013} where also a short JJ was studied by
another method (tunnel Hamiltonian method and subsequent averaging via the
Dorokhov's procedure).\cite{Dorokhov_1984}

Lempitski analyzed non-stationary behavior of long (${E_{\text{Th}} \ll
\Delta}$) S\nobreakdash-n\nobreakdash-S junctions and has shown that, in
this case, inelastic scattering rate also plays an essential role.\cite%
{Lempitskii_1983} Such ac properties of S\nobreakdash-n\nobreakdash-S JJs as
fluctuations of voltage and impedance at currents less than the critical one
were analyzed in Ref.~\onlinecite{Nagaev_1988}. The admittance of long S%
\nobreakdash-n\nobreakdash-S junctions in the frequency range ${\Omega \ll
E_{\text{Th}} / \hbar}$ has been calculated in recent papers,\cite%
{Virtanen_et_al_2011,Tikhonov_Feigelman_2015} where an expression for~$%
Y(\Omega)$ similar to Eq.~(\ref{I1}) has been obtained. This equation shows
an anomalous behavior of the admittance at low frequencies where the maximal
value of the admittance is determined by the energy relaxation rate~$\nu_{%
\text{in}}$.

In the current paper, we calculate and analyze the admittance of short JJs
of two configurations. In Section~\ref{sec:basic_eqns}, we present basic
equations for quasiclassical Green's functions which will be used in Section~%
\ref{sec:sc_wire}, where we consider a superconducting wire or film in which
the superconducting order parameter~$\Psi$ is suppressed locally so that the
amplitude~$|\Psi(x)|$ has a dip at ${x = 0}$. At strong suppression, one can
speak of a weak link. This model has much in common with the so-called
phase-slip centers\cite{Ivlev_Kopnin_1984,Arutyunov_et_al_2008} or FFLO
state in superconductors.\cite{Fulde_Ferrell_1964,Larkin_Ovchinnikov_1965}
Far away from the weak point ${x = 0}$, the ac condensate momentum~$%
Q_{\Omega}$ is connected with an ac field $E_{\Omega}$ via Eq.~(\ref{I0}).
Near this point, the momentum~$Q_{\Omega}$ depends on coordinate,~$%
Q_{\Omega}(x)$, and the gauge-invariant potential~$\mu_{\Omega}(x)$ related
to electron-hole branch imbalance arises.\cite%
{Tinkham_Clarke_1972,Tinkham_1972,Schmid75,ArtVolkovRev80} In this case, the
electric field is determined both by the gauge-invariant vector~$\mathbf{Q}$
and by the gradient of the potential~$\nabla \mu_{\Omega}(x)$ (see, for
example, Refs.~\onlinecite{Aronov_Gurevich_1974,ArtVolkovRev80}),
\begin{equation}
\hbar \frac{\partial \mathbf{Q}}{\partial t} = e \mathbf{E} + \nabla \mu \,.
\label{I1'}
\end{equation}
The gauge-invariant quantities~$\mathbf{Q}$ and~$\mu$ are defined in terms
of the vector potential~$\mathbf{A}$ and scalar electric potential~$V$
\begin{align}
\mathbf{Q} &= \frac{1}{2}(\nabla \chi - 2 \pi \mathbf{A} / \Phi_{0}) \,,
\label{I2} \\
\mu &= \frac{1}{2}(\hbar \frac{\partial \chi}{\partial t} + 2 e V) \,,
\label{I2'}
\end{align}%
where~$\chi$ is the phase of the order parameter and ${\Phi_{0} = h c / 2 e}$
is the magnetic flux quantum. Substituting Eqs.~(\ref{I2}) and~(\ref{I2'})
into Eq.~(\ref{I1'}), we obtain the standard definition of the electric
field~$\mathbf{E}$ in terms of potentials~$\mathbf{A}$ and~$V$, ${\mathbf{E}
= -(1/c) \partial_t \mathbf{A} - \nabla V}$.

On the basis of time-dependent Ginzburg-Landau equations derived by Gor'kov
and Eliashberg for gapless superconductors,\cite{G-Eliash} we find both
quantities~$Q_{\Omega}(x)$ and~$\mu_{\Omega}(x)$, and calculate the
admittance of the system. We will show that the last term at the right is
comparable with the first one and therefore can not be neglected as it was
done in some papers.

In Section~\ref{sec:ScS} we consider a short S\nobreakdash-c\nobreakdash-S
contact. By using a rather general formula for admittance derived in Ref.~%
\onlinecite{AVZ79}, we analyze the admittance of this~JJ. [The authors of
Ref.~\onlinecite{AVZ79} provided the expression Eq.~(\ref{I1}) without
considering arbitrary frequencies and temperatures.] In this case, the
electric field~$E_{\Omega}$ is connected with the phase difference~$%
\varphi_{\Omega}$ in superconducting reservoirs~S which are assumed to be in
equilibrium. We present the dependence~$Y(\Omega)$ for different values of
constant phase difference~$2 \varphi_{0}$ and arbitrary frequencies. We show
that an interesting peculiarity in this dependence arises near the point ${%
\varphi_{0} \simeq \pi / 4}$ corresponding to the critical current~$I_{\text{%
c}}$. Whereas the real part of admittance ${Y^{\prime}(\Omega) = \mathrm{Re}%
[Y(\Omega)]}$ increases smoothly with increasing~$\Omega$ at ${2 \varphi_{0}
< \pi / 2}$, it has a maximum if the phase difference~$2 \varphi_{0}$
exceeds~$\pi / 2$. Although the latter case corresponds to unstable points
on the curve~$I_{\text{J}}(\varphi_{0})$ in current-biased JJs, it can be
realized in phase-biased JJs making the predicted effect observable.\cite%
{Dassonneville_et_al_2013,Ronzani_et_al_2016_arxiv}

In the Conclusion, we discuss the possibilities to study ac properties of
the considered JJs experimentally. Note that a hump in the real part of
admittance~$Y^{\prime}(\Omega)$ at high temperatures and low~$\Omega$ is
much broader than the peak in~$Y^{\prime}(\Omega)$ caused by a resonance
excitation of the AM in uniform superconductors and is due to another
mechanism.\cite{Moor_Volkov_Efetov_2017_PRL}

\section{Basic Equations}

\label{sec:basic_eqns}

In this Section, we present basic equations for quasiclassical Green's
functions including the Keldysh function which is needed in a non-stationary
case. These equations were employed in our previous work for analysis of a
uniform case\cite{Moor_Volkov_Efetov_2017_PRL} and will be used for
calculating the admittance of a non-uniform superconductor, i.e., a short S%
\nobreakdash-c\nobreakdash-S~JJ. We have shown earlier that the AM can be
excited even by a weak ac field~$\mathbf{E}(t)$ in the presence of a
condensate flow. In addition, it was shown that the resonance excitation of
the AM contributes to the admittance~$Y(\Omega)$ of such a superconductor.
Unlike the experiments in terahertz frequency region,\cite%
{ExpS13,Matsunaga_et_al_2014} the absorption of microwave ac field in
superconductors was measured long ago by Martin and Tinkham\cite{Martin_Tinkham_1968} and later on by Budzinski~\emph{et al.}\cite{Budzinski_Garfunkel_Markley_1973} It was found that a peak near the
frequency ${\Omega = 2 \Delta / \hbar}$ arises by applying a magnetic field.
The formula describing correctly this peak was obtained by the method of
analytical continuation in Ref.~\onlinecite{Ovchinnikov_Isaakyan_1978}, the
authors of which explained the maximum in the absorption with a singularity
in the density of states but did not relate it with the resonance excitation
of the amplitude (Higgs) mode.

Like in Ref.~\onlinecite{Moor_Volkov_Efetov_2017_PRL}, we consider the
diffusive limit in one-dimensional geometry so that ${\mathbf{Q} = (Q,0,0)}$%
. The current~$\mathbf{I}_{\Omega}$ and the gap perturbation~$\delta
\Delta_{\Omega}$ are found from nonstationary equations for matrix
quasiclassical Green's functions~$\check{g}$. These equations, in the
absence of a magnetic field, have the form\cite%
{Usadel,Kopnin,LO,RammerSmith,BelzigRev,ArtVolkovRev80}
\begin{equation}
-i D \partial_{x}(\check{g} \partial_{x} \check{g}) + i(\check{\tau}_{3}
\cdot \partial_{t} \check{g} + \partial_{t^{\prime}} \check{g} \cdot \check{%
\tau}_{3}) + [ \check{\Sigma} \,, \check{g} ] = V(t) \check{g} - \check{g}
V(t^{\prime}) \,.  \label{U1}
\end{equation}
The diagonal matrix elements of the matrix~$\check{g}$ are the retarded
(advanced) Green's functions~$\hat{g}^{R(A)}$, and the off-diagonal element
is the Keldysh function~$\hat{g}$,
\begin{equation}
\check{g} =
\begin{pmatrix}
\hat{g}^{R} & \hat{g} \\
0 & \hat{g}^{A}%
\end{pmatrix}
\,.  \label{U2}
\end{equation}

The functions~$\hat{g}^{R(A)}$ and~$\hat{g}$ are $2 \times 2$~matrices in
the particle-hole space. All the functions depend on two times~$t$ and~$%
t^{\prime}$. The diagonal matrix~$\check{\Sigma}$ consists of matrices ${%
\hat{\Sigma}^{R(A)} = \Delta i \hat{\tau}_{2} + i \hat{\gamma}^{R(A)}}$,
where $\Delta$ is the superconducting gap and~$\hat{\gamma}$ is a damping
matrix. The matrix~$\check{g}(t,t^{\prime})$ obeys the normalization
condition
\begin{equation}
\check{g} \cdot \check{g} \equiv \int dt_{1} \check{g}(t, t_{1}) \cdot
\check{g}(t_{1}, t^{\prime}) = \check{1} \delta (t-t^{\prime}) \,.
\label{U3}
\end{equation}

The current in the diffusive limit is determined by the expression
\begin{equation}
\mathbf{I}(t) = - \frac{\pi \sigma}{4 e} \int dt_{1} \mathrm{Tr} \big\{
\check{\tau}_{3} \check{g}(t, t_{1}) \nabla \check{g}(t_{1}, t) \big\}^K \,,
\label{U4}
\end{equation}
where~$\sigma$ is the conductivity.

In equilibrium and in absence of a dc current, the Green's functions~$\hat{g}%
^{R(A)}$ and~$\hat{g}$ have the form
\begin{align}
\hat{g}_{\text{eq}}^{R(A)} &= [ g \hat{\tau}_{3} + f i \hat{\tau}_{2}
]^{R(A)} \,,  \label{U5} \\
\hat{g}_{\text{eq}} &= (\hat{g}^{R} - \hat{g}^{A}) \tanh (\epsilon \beta )
\,,  \label{U5'}
\end{align}
where ${g_{\text{eq}}^{R(A)} = (\epsilon / \Delta) f_{\text{eq}}^{R(A)} =
\epsilon / \zeta^{R(A)}(\epsilon)}$, ${\beta = 1 / 2 T}$ and
\begin{equation}
\zeta^{R(A)}(\epsilon) = \sqrt{(\epsilon \pm i\gamma)^{2} - \Delta^{2}} \,.
\label{U6}
\end{equation}%
The matrices~$\hat{\tau}_{i}$ are the Pauli matrices operating in the
particle-hole space.

We calculate the impedance and the gauge-invariant quantities~$Q$ and~$\mu$
in the next Section where nonstationary Ginzburg-Landau equations\cite%
{G-Eliash} will be used instead of more complicated Eqs.~(\ref{U1})--(\ref%
{U3}).

\section{Superconducting wire with a local gap suppression}

\label{sec:sc_wire}

\begin{figure}[tbp]
\includegraphics[width=1.0\columnwidth]{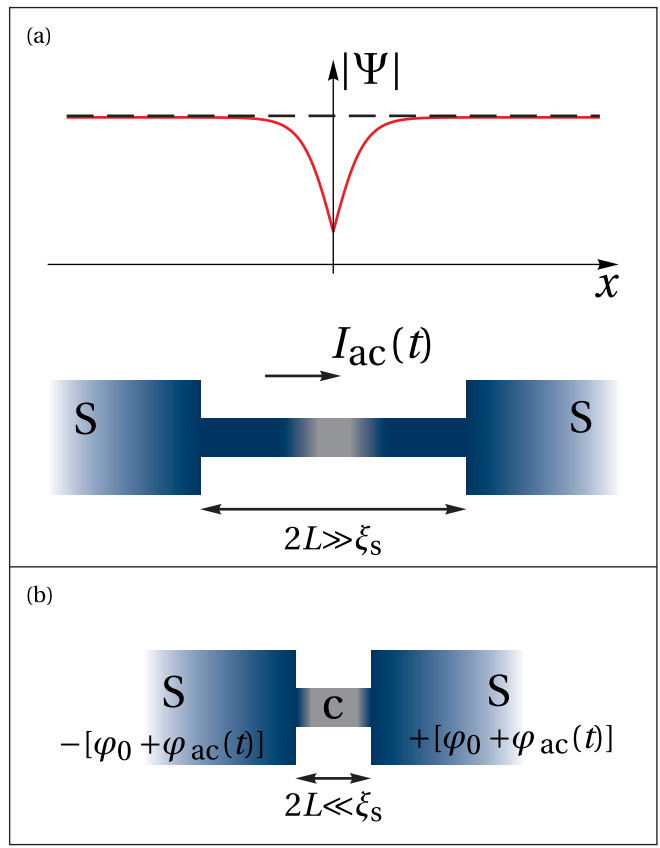}
\caption{(Color online.) Schematic view of the system under consideration.
(a)~The general suppression of the order parameter in a junction and a
general setup of a weak link of the length~${2 L \gg \protect\xi_{\text{s}}}$%
; (b)~a short weak link (${2 L \ll \protect\xi_{\text{s}}}$) with
corresponding phases of the order parameters in the superconductors forming
the junction.}
\label{fig:System}
\end{figure}

We consider a one-dimensional superconducting wire or film in which the
superconducting order parameter~${\Delta \propto \Psi}$ is locally
suppressed, see Fig.~\ref{fig:System}~(a). Our aim is to calculate the
impedance (or admittance) of this system. We describe the system under
consideration on the basis of non-stationary Ginzburg-Landau equations that
have been derived by Gor'kov and Eliashberg\cite{G-Eliash} and were used in
many papers. These equations are valid for gapless superconductors with a
high concentration of paramagnetic impurities. In the normalized form they
have the form
\begin{align}
\partial_t f &= \partial_{xx}^{2} f + f [a(x) - f^{2}] - Q^{2} f \,,
\label{1} \\
\nu f^{2} \mu &= - \partial_{x} E \,,  \label{2} \\
I &= Q f^{2} + E \,,  \label{3} \\
\partial_t Q &= E + \partial_{x} \mu \,.  \label{4}
\end{align}
Here, ${f = |\Psi| / |\Psi_{\infty}|}$ is the dimensionless modulus of the
order parameter~$\Psi$, where ${|\Psi _{\infty}| = \pi \sqrt{(T_{\text{c}%
}^{2} - T^{2})}}$. The length and time are measured in the units ${\xi_{%
\text{sf}} = \sqrt{12 D t_{0}}}$ and ${t_{0} = \hbar^{2}/(2 \tau_{\text{sf}}
\Delta^{2})}$, where~$\tau_{\text{sf}}$ is the spin-flip relaxation time.
The current~$I$ and the voltage~$V$ are measured in units of~$\sigma V_{0} /
\xi_{\text{sf}}$ and~${V_{0} = \hbar / 2 e t_{0}}$. The gauge-invariant
quantities~$\mathbf{Q}$ and~$\mu$ are defined in Eqs.~(\ref{I2}) and~(\ref%
{I2'}).

The magnitude of the relaxation rate~$\nu$ of the normalized potential~$\mu$
depends on the choice of the model. In the model of a gapless superconductor
with paramagnetic impurities considered in Ref.~\onlinecite{G-Eliash} ${\nu
= 12}$. The value of~$\nu$ in conventional BCS superconductors is much
smaller.\cite{ArtVolkovRev80} The coefficient~$a(x)$ describes a suppression
of~$|\Psi|$, respectively,~$f$. We consider the simplest model when~$a(x)$
has the form
\begin{equation}
a(x) = 1 - a_{0} \delta(x) \,,  \label{5}
\end{equation}
where the parameter~$a_{0}$ can be either small (weak suppression of~$f$) or
large (strong suppression of~$f$). The reasons for the suppression of~$%
\Delta $ can be different. For example, a locally enhanced concentration of
paramagnetic impurities leads to such a suppression. Note that the
stationary and non-stationary Josephson effects for large~$a_{0}$ have been
studied in Ref.~\onlinecite{Volkov71,*Volkov74}.

From Eq.~(\ref{5}) we find the matching condition
\begin{equation}
2 \partial_{x} f(x)|_{x=0} = a_{0} f(0) \,.  \label{6}
\end{equation}

In this Section, we consider the case when only ac current flows through the
system. From Eqs.~(\ref{1})--(\ref{5}), one needs to find a spatial
dependence~$f_{0}(x)$ in a stationary case and then to determine the linear
response to the ac current~$I_{\text{ac}}(t)$ in the system. Consider first
the stationary case.

\subsection{Stationary case}

\label{sec:stat_case}

In absence of a constant current (${I_{0} = 0}$) we need to find a
stationary solution only for Eq.~(\ref{1}) complemented by the boundary
condition Eq.~(\ref{6}) because the functions~$Q_{0}$,~$\mu$ and~$E$ vanish.
The solution is
\begin{equation}
f_{0}(x) = \tanh X  \label{7}
\end{equation}
with ${X = \kappa_{0} (|x| + x_{0})}$ and ${\kappa_{0}^{2} = 1/2}$. The
integration constant ${X_{0} \equiv \kappa_{0} x_{0}}$ is found from the
matching condition Eq.~(\ref{6}),
\begin{equation}
\sinh (2 X_{0}) = \frac{4 \kappa_{0}}{a_{0}} \,.  \label{8}
\end{equation}

In the case of weak (${a_{0} \ll 1}$), respectively, strong (${a_{0} \gg 1}$%
) suppression, the constant ${X_{0} = \kappa_{0} x_{0}}$ is
\begin{equation}
X_{0} =
\begin{cases}
2^{-1} \ln (8 \kappa_{0} / a_{0}) \,, & a_{0} \ll 1 \,, \\
2 \kappa_{0} / a_{0} \,, & a_{0} \gg 1 \,.%
\end{cases}
\label{8'}
\end{equation}
The dependence~$f_{0}(x)$ is shown schematically in Fig.~\ref{fig:System}%
~(a).

Next, we consider the non-stationary case.

\subsection{Non-stationary case}

\label{sec:nonstat_case}

Having determined the stationary function~$f_{0}(x)$, we can find the linear
response, i.e., the functions~$Q_{\Omega}$,~$\mu_{\Omega}$ and~$E_{\Omega}$
in the presence of a weak ac current
\begin{equation}
I_{\text{ac}}(t) = I_{\Omega} \cos (\Omega t) \,.  \label{10}
\end{equation}

We can linearize Eqs.~(\ref{1})--(\ref{4}). Far away from the point ${x = 0}$%
, where the normalized order parameter ${f(x) \to 1}$, we obtain
\begin{align}
E_{\infty} = \frac{-i \Omega}{1 - i \Omega} I_{\Omega} \,, \\
Q_{\infty} = \frac{1}{1 - i \Omega} I_{\Omega} \,.  \label{11}
\end{align}

Deviations from these values, ${\delta E_{\Omega} = E_{\Omega} - E_{\infty}}$
and ${\delta Q_{\Omega} = Q_{\Omega} - Q_{\infty}}$, arise due to a local
suppression of superconductivity at ${x = 0}$. We introduce a function~$%
\mathcal{E}_{\Omega}(x)$ which is connected with~$\delta E_{\Omega}(x)$ via
the relation ${\delta E_{\Omega} = f_{0}(x) \mathcal{E}_{\Omega}}$. The
function~$\mathcal{E}_{\Omega }$ obeys the equation (see Appendix~\ref%
{app:GL_Eq})
\begin{equation}
\Big[-\partial_{xx}^{2} + \nu (\tanh^{2} X - i \Omega) + \frac{1}{\sinh^{2} X%
}\Big] \mathcal{E}_{\Omega} = \frac{- 2 i I_{\Omega} \Omega \nu }{(1 - i
\Omega) \sinh (2X)} \,.  \label{16}
\end{equation}%
The boundary condition at ${x=0}$ for the function~$\mathcal{E}_{\Omega }$
is
\begin{equation}
2 \partial_{x} \mathcal{E}_{\Omega}|_{0+} = -a_{0} \mathcal{E}_{\Omega}(0)
\,.  \label{BC}
\end{equation}

We need to solve Eq.~(\ref{16}) and to find an even function~$\mathcal{E}%
_{\Omega}(x)$ decaying to zero at ${x \to \infty}$. The ac voltage~$\delta
V_{\Omega}$ across the junction is expressed through~$\mathcal{E}_{\Omega}$
via
\begin{equation}
\delta V_{\Omega} = 2 \int_{0}^{\infty} dx \, f_{0}(x) \mathcal{E}%
_{\Omega}(x) \,.  \label{17}
\end{equation}

The complex impedance of the system consists of two parts ${Z_{\Omega} =
Z_{\Omega L} + \delta Z_{\Omega}}$, where the first term is the impedance in
the absence of the weak link (${a_{0} = 0}$) and the second term is related
to the presence of the local suppression
\begin{align}
Z_{\Omega L} &= \frac{-i \Omega}{1 - i \Omega} 2 L \,, \\
\delta Z_{\Omega} &= \frac{\delta V_{\Omega}}{I_{\Omega}} \,.  \label{18a}
\end{align}
Note that for small~$a_{0}$ the problem can be solved analytically. Consider
first this case.

\subsubsection{Weak local suppression}

As follows from Eq.~(\ref{8}), for small~$a_{0}$ we have ${\sinh X \simeq
\exp[(|x| + x_{0}) / \sqrt{2}] \gg 1}$. In the main approximation, Eq.~(\ref%
{16}) can be written in the form
\begin{equation}
-\partial_{xx}^{2} \mathcal{E}_{\Omega} + \mathcal{E}_{\Omega}
\kappa_{\Omega}^{2} = - 4 i \Omega I_{\Omega} \frac{\nu \exp[-\sqrt{2}%
(|x|+x_{0})]}{1 - i \Omega} \,,  \label{19}
\end{equation}
where ${\kappa_{\Omega}^{2} = \nu (1 - i \Omega)}$. In the case of a small
parameter~$a_{0}$, a solution with continuous functions~$\mathcal{E}%
_{\Omega}(x)$ and $\partial_{x} \mathcal{E}_{\Omega}(x)$ is
\begin{align}
\mathcal{E}_{\Omega}(x) &= \frac{- 4 i a_{0} \Omega \nu I_{\Omega}}{(1 - i
\Omega)(\kappa_{\Omega}^{2} - 2)}  \label{20} \\
&\times \Big[-\frac{\sqrt{2}}{\kappa_{\Omega}} \exp (-\kappa_{\Omega} |x|) +
\exp(-\sqrt{2} |x|) \Big] \exp (- \sqrt{2} |x_{0}|) \,.  \notag
\end{align}

For the voltage~$\delta V_{\Omega}$ and the impedance~$\delta Z_{\Omega}$ we
obtain
\begin{align}
\delta V_{\Omega} &= a_{0} \frac{-i \Omega I_{\Omega}}{(1 - i \Omega)^{2}} \\
\intertext{and}
\delta Z_{\Omega} &= a_{0} \frac{-i \Omega}{(1 - i \Omega)^{2}} \,,
\label{21}
\end{align}
respectively. Therefore, the impedance variation ${\delta Z_{\Omega} =
\delta Z_{\Omega}^{\prime} + i \delta Z_{\Omega}^{\prime \prime}}$ is given
by
\begin{align}
\delta Z_{\Omega}^{\prime} \equiv \delta R(\Omega) &= a_{0} \frac{2
\Omega^{2}}{(1 + \Omega^{2})^{2}} \,,  \label{22} \\
\delta Z_{\Omega}^{\prime \prime} &= - a_{0} \frac{\Omega (1 - \Omega^{2})}{%
(1 + \Omega^{2})^{2}} \,.  \label{23}
\end{align}
The total resistance and the reactive part of the impedance of the wire is
\begin{align}
R(\Omega) \equiv Z^{\prime}(\Omega) &= \frac{\Omega^{2}}{(1 + \Omega^{2})} %
\Big[2 L + \frac{2a_{0}}{(1 + \Omega^{2})}\Big] \,,  \label{24} \\
Z^{\prime \prime}(\Omega) &= - \frac{\Omega}{(1 + \Omega^{2})} \Big[2 L +
\frac{a_{0} (1 - \Omega^{2})}{(1 + \Omega^{2})}\Big] \,.
\end{align}
One can see that the active part of the impedance increases due to a
suppression of the order parameter~$f$ at ${x = 0}$. The reactive part
increases at ${\Omega \leq 1}$ and decreases at ${\Omega \geq 1}$, that is,
the variation of the reactive part~$\delta Z_{\Omega}^{\prime \prime}$
changes sign at ${\Omega = 1}$.

It is of interest to find also the admittance~$Y(\Omega) \equiv 1/Z(\Omega )$%
. From Eqs.~(\ref{18a}), (\ref{22}), and~(\ref{23}) in the main
approximation in the parameter~$a_{0}$, we obtain
\begin{equation}
Y(\Omega) = \frac{1}{2L} \Big[ 1 - \frac{1 - a_{0} / 2L}{i \Omega} \Big] \,.
\label{24a}
\end{equation}
This expression shows that the considered system can be modelled as a
conductance and an inductance connected in parallel. The small gap
suppression causes a small increase in the inductance ${\mathcal{L} = 2 L /
( 1 - a_{0} / 2 L )}$ and does not change the real part of the conductance.

\subsubsection{Strong local suppression}

\begin{figure}[tbp]
\includegraphics[width=1.0\columnwidth]{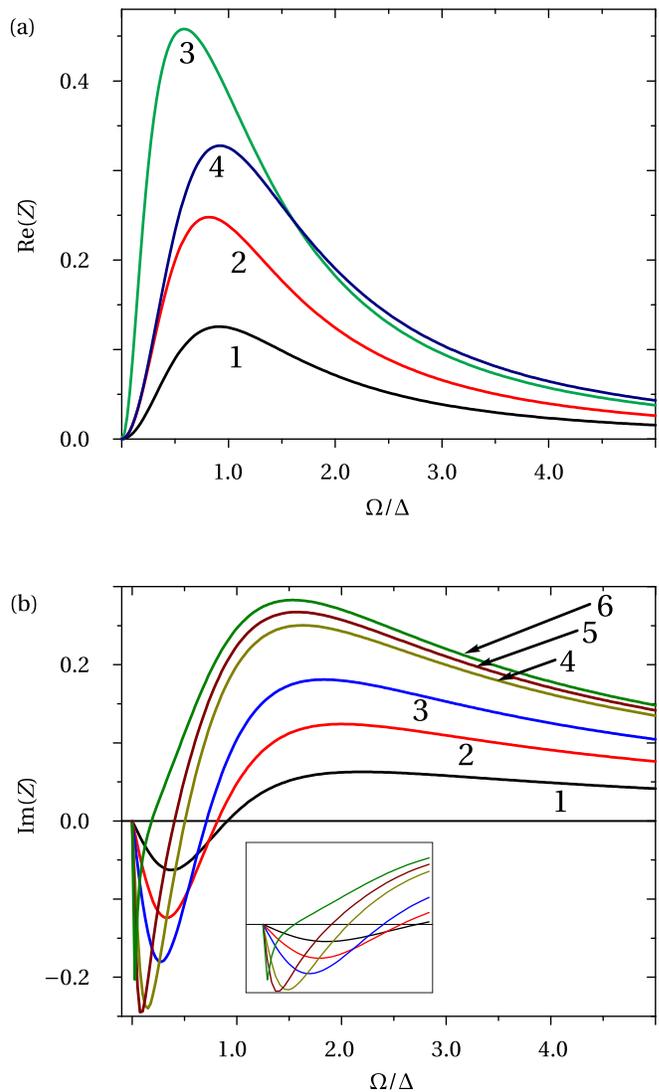}
\caption{(Color online.) (a)~Frequency dependence of the resistance
variation due to a superconductivity suppression. The numbers on the curves
denote, correspondingly, the values for~$a_0$, i.e., (1)~${a_0 = 0.5}$; (2)~$%
{a_0 = 1.0}$; (3)~${a_0 = 2.0}$; (4)~${a_0 = 5.0}$. (b)~Frequency dependence
of the variation of the reactive part of impedance due to a
superconductivity suppression. The numbers on the curves denote,
correspondingly, the values for~$a_0$, i.e., (1)~${a_0 = 0.5}$; (2)~${a_0 =
1.0}$; (3)~${a_0 = 2.2}$; (4)~${a_0 = 2.4}$; (5)~${a_0 = 2.6}$; (6)~${a_0 =
2.8}$. The inset shows the enlarged part of~$\mathrm{Im}(Z)$ at $\Omega \leq \Delta$.}
\label{fig:z}
\end{figure}

At strong suppression (${a_{0} \gg 1}$), the solution of Eq.~(\ref{16}),
which looks like the ``Schroedinger'' equation with a complex potential, can
be found numerically. In Figs.~\ref{fig:z}~(a) and~\ref{fig:z}~(b) we plot
the frequency dependence of the changes in the real and imaginary parts of
the impedance~$\delta Z_{\Omega}^{\prime}$ and~$\delta Z_{\Omega}^{\prime
\prime}$ for different values of~$a_{0}$. For small~$a_{0}$, the results of
numerical calculations and the analytical expressions given by Eqs.~(\ref{22}%
) and~(\ref{23}) coincide.

We see that the resistance due to the weak link~$\delta R(\Omega)$ is
positive and has a broad maximum at frequencies~$\Omega_{m}$ that are
slightly less than~$1$ (at small~$a_{0}$). The position of the maximum
shifts towards smaller~$\Omega_{m}$ with increasing~$a_{0}$ (when~$a_{0}$
remains less than~$\approx 2.5$). The reactive part of the impedance~$\delta
Z_{\Omega}^{\prime \prime}$ changes sign at approximately the same
frequencies. At~${a_{0} \gtrsim 2.5}$, the maximum value of~$\delta
R(\Omega) $ decreases with further increase of~$a_{0}$, whereas the
frequency~$\Omega_{m}$ increases [see Fig.~\ref{fig:z}~(a)]. The behaviour
of the reactive part~$\delta Z_{\Omega}^{\prime \prime}$ also changes. It is
worth noting that in the considered model of a gapless superconductor, the
parameter~$|\Psi|$ is the amplitude of the superconducting order parameter,
but not the gap.

\begin{figure}[tbp]
\includegraphics[width=1.0\columnwidth]{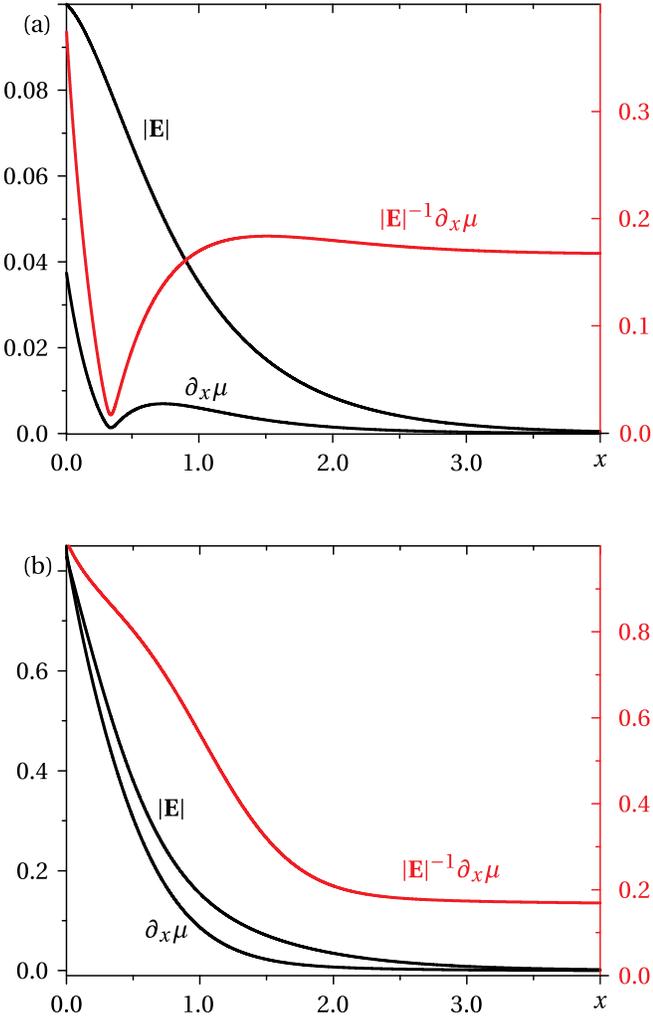}
\caption{(Color online.) Frequency dependence of the electric field
variation ${\protect\delta E_{\Omega}(x) = f_{0}(x) \mathcal{E}_{\Omega}(x)}$
and the derivative~$\partial_{x} \protect\mu_{\Omega}(x)$ as well as their
ratio for (a)~${a_{0}=0.5}$; (b)~${a_{0}=2.4}$. In both cases, the frequency
${\Omega = 0.8 \Delta}$. All quantities are corespondingly normalized, see
Eqs.~(\ref{1})--(\ref{4}).}
\label{fig:E_and_mu}
\end{figure}

\bigskip

In Figs.~\ref{fig:E_and_mu}~(a) and~\ref{fig:E_and_mu}~(b) we display the
spatial dependence of the dimensionless electric field~${\delta E_{\Omega
}(x)=f_{0}(x)\mathcal{E}_{\Omega }(x)}$ and compare it to the magnitude of
the spatial derivative of the gauge-invariant potential~$\partial _{x}\mu
_{\Omega }(x)$ for two values of~$a_{0}$, i.e., for ${a_{0}=0.5}$ (weak
suppression) and ${a_{0}=2.4}$ (strong suppression). One can see that these
quantities may be comparable in their values. This means that the electric
field~$\mathbf{E}_{\Omega }$ is not determined only by the condensate
momentum~$\mathbf{Q}_{\Omega }$ [see Eq.~(\ref{I0}) which is valid in a
uniform case] and that in order to find the linear response of a
superconductor with a non-homogeneous order parameter~$f(x)$, the potential~$%
\mu _{\Omega }(x)$ has to be calculated also alonside with $\mathbf{Q}%
_{\Omega }$ if the ac electric field~$\mathbf{E}_{\Omega }$ is directed
parallel to $x$~axis. This statement is true, for instance, for the case of
the FFLO state (compare it with Ref.~\onlinecite{Dai_Lee_2017}, where the
optical conductance of a non-homogeneous superconductor was calculated in
the gauge with ${\mathbf{A}\neq 0}$ and ${V=0}$ so that ${\mu =0}$).

\section{S-c-S contact}

\label{sec:ScS}

In this Section, we consider short Josephson junctions of the S\nobreakdash-c%
\nobreakdash-S or S\nobreakdash-n\nobreakdash-S types in the dirty limit,
i.e., in the limit ${\tau T_{\text{c}} \ll 1}$, where~$\tau$ is the momentum
relaxation time. We also assume that there are no barriers at the S%
\nobreakdash-c interfaces. In the considered model, two superconducting
reservoirs~S are connected by a narrow constriction. Since the length of the
constriction~$2L$ is assumed to be less than the coherence length ${\xi_{%
\text{S}} \simeq \sqrt{D/T_{\text{c}}}}$, that is, the Thouless energy is
large (${D/L^{2} \gg T_{\text{c}}}$), it does not matter whether the
constriction is normal or superconducting.

Formula for the impedance~$Z(\Omega )$ in this case has been obtained by one
of the authors (in collaboration with Artemenko and Zaitsev) in~1979\cite%
{AVZ79} on the basis of microscopic theory of the Josephson effects in these
JJs, but it has not been analyzed in detail. Here, we reproduce the main
steps of the derivation of this expression, correct typos in Ref.~%
\onlinecite{AVZ79} and analyze the admittance of the short S\nobreakdash-c\nobreakdash-S JJs
in more detail. [The signs in Eq.~(29) of Ref.~\onlinecite{AVZ79} should be changed in such a way that expressions in the curly brackets in Eqs.~(27) and~(29) coincide with each other if the functions $g^{A}(\epsilon_{-})$ in Eq.~(27) are replaced by~$g^{R}(\epsilon_{-})$. The imaginary unit~$i$ in front of the right-hand side of Eq.~(29) has to be dropped. The last term in Eq.~(31) should have the form~$\hbar i \omega \varphi_{\omega} / 2 e R$. Note that~$\omega$ in Ref.~\onlinecite{AVZ79} corresponds to~$-\Omega$.] Note that the admittance of a similar S\nobreakdash-c%
\nobreakdash-S contact has been calculated and analyzed in a recent paper,%
\cite{Kos_Nigg_Glazman_2013} where another model and method of calculations
were used.

The microscopic theory developed in Ref.~\onlinecite{AVZ79} is based on the
generalized Usadel equation, Eq.~(\ref{U1}), which describes the spatial
dependence of the Green's functions~$\check{g}(x, t, t^{\prime})$ in the
constriction. These functions are assumed to be continuous at the S%
\nobreakdash-c and c\nobreakdash-S~interfaces (no potential barriers at
these interfaces).

In the considered limit of a short junction, one can neglect all the terms
in Eq.~(\ref{U1}) except the first one and we obtain for the ``anisotropic''
part
\begin{equation}
\check{a} = -l \check{g} \partial_{x} \check{g} = \text{const} \,.
\label{25}
\end{equation}
That is, the matrix~$\check{a}$ does not depend on the coordinate~$x$. The
current~$I$ through the considered JJ is expressed through the anisotropic
part of the Keldysh function~$\hat{a}$ as follows
\begin{equation}
I(t) = - \frac{\pi \sigma}{4 e} \int d t_1 \mathrm{Tr}\big\{ \hat{\tau}_{3}
\hat{g} \partial_x \hat{g} \big\}^K \,.  \label{26}
\end{equation}

A formal solution of Eq.~(\ref{25}) is (for brevity we drop the temporal
indices~$t$ and~$t^{\prime}$)
\begin{equation}
\check{g}(x) = \check{g}(0) \exp (-\check{a} x / l) \,.  \label{27}
\end{equation}

As follows from Eq.~(\ref{U3}), the matrices~$\check{a}$ and~$\check{g}(0)$
anticommute and ${\check{g}(0) \cdot \check{g}(0) = \check{1}}$. Thus,
introducing the matrices ${\check{G}^{(\pm)} \equiv \big[ \check{G}(L) \pm
\check{G}(-L) \big] / 2}$ and using Eq.~(\ref{27}), we obtain
\begin{align}
\check{G}^{(+)} &= \check{g}(0) \cosh (\check{a} L/l) \,, \\
\check{G}^{(-)} &= \check{g}(0) \sinh (\check{a} L/l) \,,  \label{28}
\end{align}
where~${\check{G}(\pm L) \equiv \check{G}^{(\pm)}}$ are the known matrix
Green's functions in the reservoirs. From this equation we find
\begin{equation}
\check{a} = -\frac{l}{2L} \mathrm{arsinh} \big[2 \check{G}^{(+)} \cdot
\check{G}^{(-)}\big] \,.  \label{29}
\end{equation}
In particular,
\begin{equation}
\hat{a}^{R(A)} = -(l/2L) \mathrm{arsinh} \big[2 \hat{G}^{(+)} \cdot \hat{G}%
^{(-)}\big]^{R(A)} \,.  \label{29'}
\end{equation}

The matrices~$[\hat{g}^{(\pm )}]^{R(A)}$ are expressed in terms of the
retarded (advanced) Green's functions in the reservoirs ${\hat{g}^{R(A)}(\pm
L) \equiv \hat{G}^{R(A)}}$ that are known and have the form
\begin{equation}
\hat{G}^{R(A)}(\pm L) = \hat{S}(t, \pm L) \cdot \hat{G}_{0}^{R(A)}(t-t^{%
\prime}) \cdot \hat{S}^{\dagger}(t^{\prime}, \pm L) \,.  \label{30}
\end{equation}
Here, we introduce the transformation matrix ${\hat{S}(t, \pm L) = \exp (i
\hat{\tau}_{3} \varphi (t, \pm L) / 2)}$ in order to take into account the
presence of the phase of the superconducting order parameter in the banks ${%
\varphi (t, \pm L) = \pm [ \varphi_{0} + \varphi_{\text{ac}}(t) ]}$. The
Green's functions~$\hat{G}_{0}^{R(A)}(\epsilon)$ in reservoirs in the
absence of phase difference coincide with the matrices~$\hat{g}_{\text{eq}%
}^{R(A)}$ defined in Eq.~(\ref{U5}).

Consider first the stationary case.

\subsection{Stationary case}

In the equilibrium case [${\varphi_{\text{ac}}(t) = 0}$], the Keldysh
function~$\hat{a}$ depends only on the time difference ${(t - t^{\prime})}$
and its Fourier component is
\begin{equation}
\hat{a}(\epsilon) = \big[ \hat{a}_{0}^{R}(\epsilon) - \hat{a}%
_{0}^{A}(\epsilon) \big] \tanh (\epsilon \beta) \,,  \label{32}
\end{equation}
where ${\beta = 1 / 2 T}$. The matrices~$\hat{a}^{R(A)}(\epsilon)$ are found
from Eqs.~(\ref{29'}) and~(\ref{30}),
\begin{equation}
\hat{a}_{0}^{R(A)}(\epsilon) = -\frac{l}{L} \big[ \Delta \cos \varphi_{0}
\hat{\tau}_{3} + \epsilon i \hat{\tau}_{2} \big] b^{R(A)}(\epsilon)
\label{33}
\end{equation}
with
\begin{equation}
b^{R}(\epsilon) = \frac{i}{\tilde{\zeta}^{R}(\epsilon)} \mathrm{arsinh} %
\Big[ \frac{\Delta \sin \varphi_{0}}{\zeta^{R}(\epsilon)} \Big]  \label{34}
\end{equation}
and ${\tilde{\zeta}^{R}(\epsilon) = \sqrt{(\epsilon + i \gamma)^{2} -
\Delta^{2} \cos^{2} \varphi_{0}}}$. The function~$\zeta^{R}(\epsilon)$ is
defined in Eq.~(\ref{U6}).

In obtaining Eq.~(\ref{33}), we used the relation
\begin{equation}
\mathrm{arsinh} \Big[ \frac{2 \Delta \tilde{\zeta}^{R}(\epsilon) \sin
\varphi_{0}}{[\zeta^{R}(\epsilon)]^{2}} \Big] = 2 \mathrm{arsinh} \Big[
\frac{\Delta \sin \varphi_{0}}{\zeta^{R}(\epsilon)} \Big] \,,  \label{36}
\end{equation}
and the expressions for~$\big\{ \hat{G}^{(+)} \big\}^{R(A)}$ and $\big\{
\hat{G}^{(-)} \big\}^{R(A)}$, which directly follow from Eq.~(\ref{30}),
\begin{align}
\big\{ \hat{G}^{(+)} \big\}^{R(A)} &= \big\{ G(\epsilon) \hat{\tau}_{3} +
F(\epsilon) i \hat{\tau}_{2} \cos \varphi_{0} \big\}^{R(A)} \,, \\
\big\{ \hat{G}^{(-)} \big\}^{R(A)} &= \big\{ F(\epsilon) i \hat{\tau}_{1}
\sin \varphi_{0} \big\}^{R(A)} \,.  \label{37}
\end{align}

Thus, the Josephson dc current~$I_{\text{J}}$ can be easily found from Eqs.~(%
\ref{26}) and~(\ref{32}). The integration over energy~$\epsilon$ can be
transformed to the summation over Matsubara frequencies ${\omega > 0}$ for
the first term in Eq.~(\ref{32}) and over negative~$\omega$ for the second
term. As a result we obtain
\begin{equation}
I_{\text{J}} = I_{\text{c}}(\varphi_{0}) \sin (2 \varphi_{0}) \,,  \label{38}
\end{equation}
where the critical current~$I_{\text{c}}$ also depends on the phase
difference~$2 \varphi_{0}$ and is determined by the expression\cite%
{Kulik_Omelyanchuk_1978}
\begin{equation}
I_{\text{c}}(\varphi_{0}) = \frac{2 \pi T}{e R} \sum_{\omega > 0} \frac{%
\Delta}{\tilde{\zeta}_{\omega}(\varphi_{0}) \sin \varphi_{0}} \arcsin \Big(%
\frac{\Delta \sin \varphi_{0}}{\zeta_{\omega}}\Big) \,,  \label{39}
\end{equation}
where ${R^{-1} = \sigma \mathcal{S} / 2 L}$ with the cross section area of
the junction~$\mathcal{S}$.

One can see that near~$T_{\text{c}}$, when ${\arcsin (\Delta \sin
\varphi_{0} / \zeta_{\omega}) \simeq \Delta \sin \varphi_{0} / \omega}$ and $%
{\tilde{\zeta}_{\omega}(\varphi_{0}) \simeq \omega}$, the critical current
does not depend on the phase difference~$2 \varphi_{0}$ and is equal to ${I_{%
\text{c}} = (\pi / 4) (\Delta^{2} / e T R)}$.\cite{Kulik_Omelyanchuk_1978}
At low temperatures, the phase dependence of the Josephson current deviates
from the sinusoidal one.

\subsection{Non-stationary case}

In this Section, we find a linear response of the system to ac phase
variation~$\varphi_{\Omega}$. To do this, we need to find a deviation of the
Keldysh component ${\delta \hat{a} = \hat{a} - \hat{a}_{0}}$ due to
variation~$\varphi_{\Omega}$. It can be written in the form
\begin{equation}
\delta \hat{a} = \delta \hat{a}^{R} \tanh (\epsilon_{-} \beta) - \tanh
(\epsilon_{+} \beta) \delta \hat{a}^{A} + \hat{a}^{\text{an}} \,.  \label{40}
\end{equation}
The first two terms represent a regular part, which is an analytical
function in the upper (lower) half-plane, and the last term is a
non-analytical ``anomalous'' part.\cite{G-Eliash,ArtVolkovRev80}

Therefore, the current~$I$ through the S\nobreakdash-c\nobreakdash-S JJ can
be written in the form
\begin{equation}
I_{\Omega }=I_{\Omega }^{\text{reg}}+I_{\Omega }^{\text{an}}\,,  \label{47}
\end{equation}%
where
\begin{equation}
I_{\Omega }^{\text{reg}}=\frac{\varphi _{\Omega }}{8eR}\int d\,\bar{\epsilon}%
[j^{R}(\epsilon _{+},\epsilon _{-})\tanh (\epsilon _{-}\beta )-\tanh
(\epsilon _{+}\beta )j^{A}(\epsilon _{+},\epsilon _{-})]  \label{48}
\end{equation}%
and
\begin{equation}
I_{\Omega }^{\text{an}}=\frac{\varphi _{\Omega }}{8eR}\int d\,\bar{\epsilon}%
j^{\text{an}}(\epsilon _{+},\epsilon _{-})[\tanh (\epsilon _{-}\beta )-\tanh
(\epsilon _{+}\beta )]\,.  \label{49}
\end{equation}%
The functions~$j^{R}$ and~$j^{\text{an}}(\epsilon )$ are determined as
follows (see Appendix~\ref{app:current})
\begin{align}
j^{R}(\epsilon _{+},\epsilon _{-})& =\frac{\big[b(\epsilon
_{+})[(G_{+}-G_{-})\epsilon _{+}-(F_{+}+F_{-})\Delta \cos ^{2}\varphi _{0}]%
\big]^{R}}{[F(\epsilon _{+})-F(\epsilon _{-})\sin \varphi _{0}]^{R}}
\label{50} \\
& +\frac{\big[b(\epsilon _{-})[(G(\epsilon _{+})-G(\epsilon _{-}))\epsilon
_{-}+(F_{+}+F_{-})\Delta \cos ^{2}\varphi _{0}]\big]^{R}}{[F(\epsilon
_{+})-F(\epsilon _{-})\sin \varphi _{0}]^{R}}\,,  \notag
\end{align}

\begin{align}
j^{\text{an}} &= \frac{b^{R}(\epsilon_{+}) [ (G_{+}^{R} - G_{-}^{A})
\epsilon_{+} - (F_{+}^{R} + F_{-}^{A}) \Delta \cos^{2} \varphi_{0} ]}{[
F^{R}(\epsilon_{+}) - F^{A}(\epsilon_{-}) \sin \varphi_{0} ]}  \label{51} \\
&+ \frac{b^{A}(\epsilon_{-}) [ (G_{+}^{R} - G_{-}^{A}) \epsilon_{-} +
(F_{+}^{R} + F_{-}^{A}) \Delta \cos^{2} \varphi_{0} ]}{[ F^{R}(\epsilon_{+})
- F^{A}(\epsilon_{-}) \sin \varphi_{0} ]} \,.  \notag
\end{align}

Equations~(\ref{47})--(\ref{51}) together with the Josephson relation (we
assume the equilibrium state in the S~reservoirs),
\begin{equation}
V_{\Omega} = (\hbar / 2 e)(-i \Omega) \varphi_{\Omega} \,,  \label{51a}
\end{equation}
determine the admittance of the system ${Y_{\Omega} = I_{\Omega} / V_{\Omega}%
}$.

One can see that at low frequencies and temperatures $\mathrm{Re}[Y(\Omega)]$
is zero. Indeed, one can represent~$\tanh (\epsilon_{\pm} \beta)$ in the
form ${\tanh (\epsilon_{\pm} \beta) \simeq \tanh (\bar{\epsilon} \beta) \pm
(\Omega / 2) \cosh^{-2} (\bar{\epsilon} \beta)}$. Taking into account that
at ${|\bar{\epsilon}| \leq \Delta \cos (\varphi_{0})}$ ${\tilde{\zeta}^{R} =
\tilde{\zeta}^{A} = i \sqrt{\tilde{\Delta}^{2} - \epsilon^{2}}}$ and ${%
\zeta^{R} = \zeta^{A} = i \sqrt{\Delta^{2} - \epsilon^{2}}}$ coincide, we
obtain that ${G^R = G^A}$ and ${F^R = F^A}$. This means that the part of the
regular ``currents'' ${-( j^{R} + j^{A} ) (\Omega / 2) \cosh^{-2} (\bar{%
\epsilon} \beta)}$ cancels the anomalous ``current''~$j^{\text{an}} (\Omega
/ 2) \cosh^{-2}(\bar{\epsilon} \beta)$. The remaining part of the regular
``current'', ${(j^{R} - j^{A}) \tanh (\bar{\epsilon} \beta)}$, contribute
only to the imaginary part of the admittance~$\mathrm{Im}[Y(\Omega)]$.

\begin{figure}[tbp]
\includegraphics[width=1.0\columnwidth]{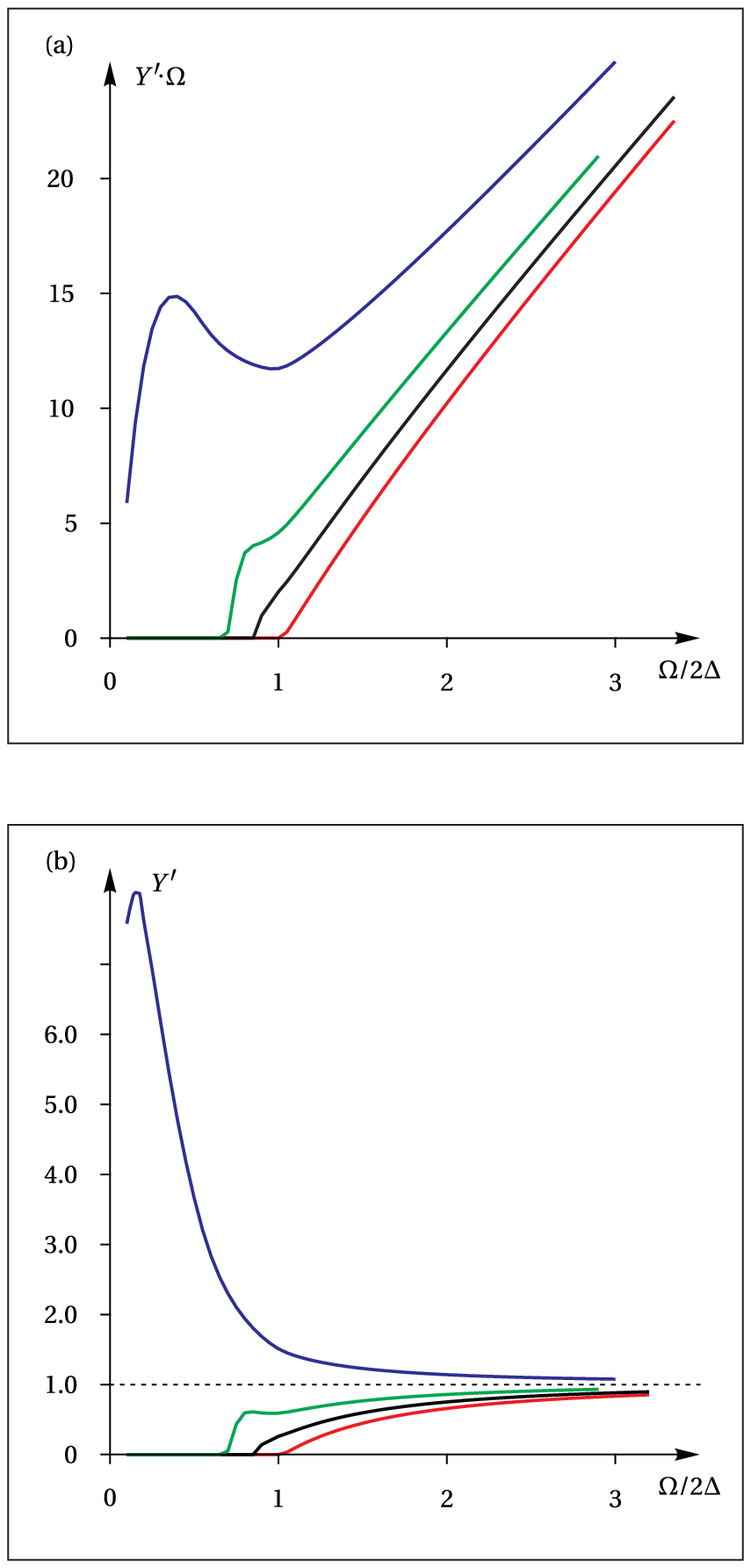}
\caption{(Color online.) (a)~Frequency dependence of the product~$Y^{\prime}(\Omega)\cdot\Omega$ which corresponds to the kernel~$Q(\Omega)$ in Fig.~8 of Ref.~\onlinecite{AbrGor58}.
(b)~Frequency dependence of the real part of
admittance~$Y^{\prime}(\Omega)$ at low temperatures for various
values of the phase difference~$2 \protect\varphi_{0}$. One can see that~$%
Y^{\prime}(\Omega)$ increases with increasing~$\Omega$ if the frequency~$%
\Omega$ exceeds a threshold value~$\Omega_{\text{Th}}$ which depends on~$%
\protect\varphi_{0}$. In the absence of the phase difference (no
supercurrent flows through the JJ) we have ${\Omega_{\text{Th}} = 2 \Delta /
\hbar}$. The curves correspond to ${\cos(2 \protect\varphi_0) = 1}$ (red), ${%
\cos(2 \protect\varphi_0) = 0.87}$ (black), ${\cos(2 \protect\varphi_0) = 1/%
\protect\sqrt{2}}$ (green), and ${\cos(2 \protect\varphi_0) = 0}$ (blue). At
low temperatures, the real part of the admittance~$Y^{\prime}(\Omega)$
increases monotonously with increasing~$\Omega$ if the latter exceeds~$2
\Delta$ and ${\protect\varphi_0 \leq \protect\varphi_{\text{c}}}$, where~$%
\protect\varphi_{\text{c}}$ is the phase difference corresponding to
critical current. At ${\protect\varphi_0 > \protect\varphi_{\text{c}}}$, the
admittance has a maximum at small~$\Omega$.}
\label{fig:Re_Y}
\end{figure}

In Fig.~\ref{fig:Re_Y}~(a) we desplayed the frequency dependence of the product~$Y^{\prime}(\Omega) \Omega$ (where $\Omega$ is normalized to~$2\Delta $) which is proportional to the kernel~$Q(\Omega)$ in Fig.~8 of Ref.~\onlinecite{AbrGor58}, where the kernel~$Q$ has been calculated for a uniform dirty superconductor.

In Fig.~\ref{fig:Re_Y}~(b), we present the frequency dependence of the real part of admittance~$Y^{\prime}(\Omega)$ (normalized to its value in the normal state) at low temperatures for
various values of the phase difference~$2\varphi _{0}$. One can see that~$%
Y^{\prime }(\Omega )$ increases with increasing~$\Omega $ if the frequency~$%
\Omega $ exceeds a threshold value~$\Omega _{\text{Th}}$ which depends on~$%
\varphi _{0}$. In the absence of the phase difference (no supercurrent flows
through the JJ) we have ${\Omega _{\text{Th}}=2\Delta /\hbar }$. The curves
correspond to ${\cos (2\varphi _{0})=1}$ (red), ${\cos (2\varphi _{0})=0.87}$
(black), ${\cos (2\varphi _{0})=1/\sqrt{2}}$ (green), and ${\cos (2\varphi
_{0})=0}$ (blue). At low temperatures, the real part of the admittance~$%
Y^{\prime }(\Omega )$ increases monotonously with increasing~$\Omega $ if
the latter exceeds~$2\Delta $ and ${\varphi _{0}\leq \varphi _{\text{c}}}$,
where~$\varphi _{\text{c}}$ is the phase difference corresponding to
critical current. At ${\varphi _{0}>\varphi _{\text{c}}}$, the admittance
has a maximum at small~$\Omega $.

As we noted in the Introduction, an interesting behavior of the admittance
takes place at low~$\Omega $ and high temperatures (${T\gg \Delta }$). The
main contribution to the real part~$Y^{\prime }(\Omega )$ [see Eq.~(\ref{47}%
)] stems from~$j^{\text{an}}$. Integration over large energies~$\epsilon $ ($%
{\epsilon \gg \Delta }$) gives the second term~$1/R$ at the right hand side
of Eq.~(\ref{I1}). In this case, ${F^{R}(\epsilon _{+})\simeq
-F^{A}(\epsilon _{-})\simeq \Delta /\epsilon }$, ${G^{R}(\epsilon
_{+})\simeq -G^{A}(\epsilon _{-})\simeq 1}$, and ${b^{R}\simeq b^{A}\simeq
i\Delta \sin \varphi _{0}/\epsilon ^{2}}$. The largest contribution occurs
due to the first terms in the square brackets in Eq.~(\ref{51}). Integrating
these terms,
\begin{align}
I_{1\Omega }^{\text{an}}& =\frac{i\varphi _{\Omega }}{4eR}\int d\,\bar{%
\epsilon}[\tanh (\epsilon _{+}\beta )-\tanh (\epsilon _{-}\beta )]
\label{52} \\
& \simeq \frac{i\Omega \varphi _{\Omega }}{2eR}\int_{0}^{\infty }d\,\bar{%
\epsilon}\cosh ^{-2}(\bar{\epsilon}\beta )=\frac{V_{\Omega }}{R}\,,  \notag
\end{align}%
we obtain the main contribution to the the admittance~$1/R$.

\begin{figure}[tbp]
\includegraphics[width=1.0\columnwidth]{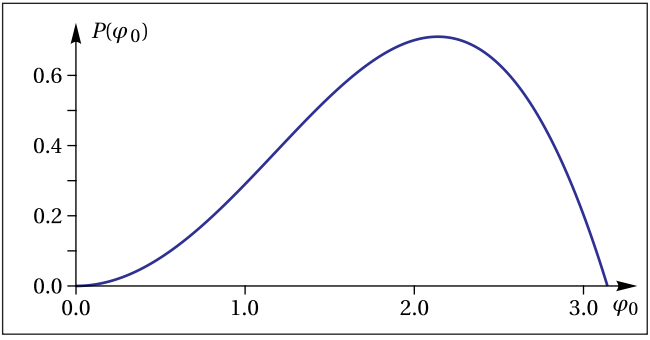}
\caption{(Color online.) The form of the function of $P(\protect\varphi_{0})$%
.}
\label{fig:P}
\end{figure}

The second important contribution to~$Y^{\prime }(\Omega )$ stems from the
second term in the square brackets at the right hand side of Eq.~(\ref{51})
in the energy interval
\begin{equation}
\tilde{\Delta}=\Delta \cos \varphi _{0}\leq \epsilon \leq \Delta \,.
\label{53'}
\end{equation}%
In this interval, we have ${F^{R}(\epsilon _{+})-F^{A}(\epsilon _{-})\simeq
(\Omega /2+i\nu _{\text{in}})\partial _{\epsilon }F^{R}}$, ${\tilde{\zeta}%
^{R}=-\tilde{\zeta}^{A}}$ and ${\mathrm{arsinh}\big(\frac{\Delta \sin
\varphi _{0}}{\zeta ^{R}}\big)+\mathrm{arsinh}\big(\frac{\Delta \sin \varphi
_{0}}{\zeta ^{A}}\big)=-i\pi }$. Therefore, setting $\cosh
^{-2}(\epsilon \beta )\approx 1$, we obtain at ${T\gg \Delta }$
\begin{align}
I_{2\Omega }^{\text{an}}& =\frac{2eV_{\Omega }}{\hbar }\frac{\pi \Delta ^{2}%
}{2T}\frac{\gamma _{\epsilon }-i\Omega }{\gamma _{\epsilon }^{2}+\Omega ^{2}}%
\frac{\cos ^{2}\varphi _{0}}{\sin \varphi _{0}}\int_{\cos \varphi
_{0}}^{1}d\,x\frac{1-x^{2}}{x\sqrt{x^{2}-\cos ^{2}\varphi _{0}}} \\
& =\frac{2eV_{\Omega }}{\hbar }I_{c}\frac{\gamma _{\epsilon }-i\Omega }{%
\gamma _{\epsilon }^{2}+\Omega ^{2}}P(\varphi _{0})\,,  \notag
\end{align}%
where ${\nu _{\text{in}}=\gamma _{\epsilon }/2}$. Thus, the admittance is
given by Eq.~(\ref{I1}) with the function~$P(\varphi _{0})$ equal to
\begin{equation}
P(\varphi _{0})=\cot (\varphi _{0})[2\varphi _{0}-\sin (2\varphi _{0})]\,.
\label{54}
\end{equation}%
The function~$P(\varphi _{0})$ is shown in Fig.~\ref{fig:P}.

The integral in Eq.~(\ref{52}) can be calculated for any temperatures if the factor~$\cosh^{-2}(\epsilon \beta)$ in the integrand is
taken into account, i.e., not using the approximation ${\cosh^{-2}(\epsilon \beta )\approx 1}$.

Therefore, the deviation~$\delta Y^{\prime }(\Omega )$ of the real part of
the admittance~$Y^{\prime }(\Omega )$ from its value in the normal state~$%
(1/R)$ is
\begin{equation}
\delta Y^{\prime }(\Omega )R=\frac{2eI_{\text{c}}R}{\hbar }\frac{\gamma
_{\epsilon }}{\gamma _{\epsilon }^{2}+\Omega ^{2}}P(\varphi _{0})\,.
\label{55}
\end{equation}

The normalized deviation ${\delta \tilde{Y}^{\prime}(\Omega) \equiv \delta
Y^{\prime}(\Omega) R}$ has a maximum at ${\Omega = 0}$ with a magnitude ${%
\delta \tilde{Y}^{\prime}(\Omega)_{\max} \simeq 2 e I_{\text{c}} R / \hbar
\nu_{\text{in}} \simeq \Delta^{2} / (T \hbar \nu_{\text{in}})}$ which can be
much larger than~1. The enhancement of the admittance, Eq.~(\ref{55}), is
caused by quasiparticles with energies in the interval defined by Eq.~(\ref%
{53'}).

It is of interest to calculate the density of states (DOS)~$N(\epsilon,x)$ in the junction and its spatial dependence. Note that this
dependence cannot be found in tunnel Hamiltonian approach. The function~$%
N(\epsilon ,x)$ is zero at energies ${|\epsilon |\leq \tilde{\Delta}}$, but
is finite at energies ${|\epsilon |\geq \tilde{\Delta}}$. In this energy
range, ${\tilde{\Delta}\leq |\epsilon |\leq \Delta }$, the DOS~$%
N_{<}(\epsilon ,x)$ is given by (see Appendix~\ref{app:DOS_in_SCS} and Ref.~%
\onlinecite{AslVolkov86})
\begin{equation}
N_{<}(\epsilon ,x)=\frac{|\epsilon |}{\sqrt{\epsilon ^{2}-\tilde{\Delta}^{2}}%
}\cosh (\tilde{x}\ln M_{<})\cos \Big(\frac{\pi }{2}\tilde{x}\Big)\,,
\label{56a}
\end{equation}%
where the function~$M_{<}$ is
\begin{equation}
M_{<}=\frac{\Delta \sin \varphi _{0}+\sqrt{\epsilon ^{2}-\Delta ^{2}\cos
^{2}\varphi _{0}}}{\sqrt{\Delta ^{2}-\epsilon ^{2}}}\,.  \label{56b}
\end{equation}

\begin{figure}[tbp]
\includegraphics[width=1.0\columnwidth]{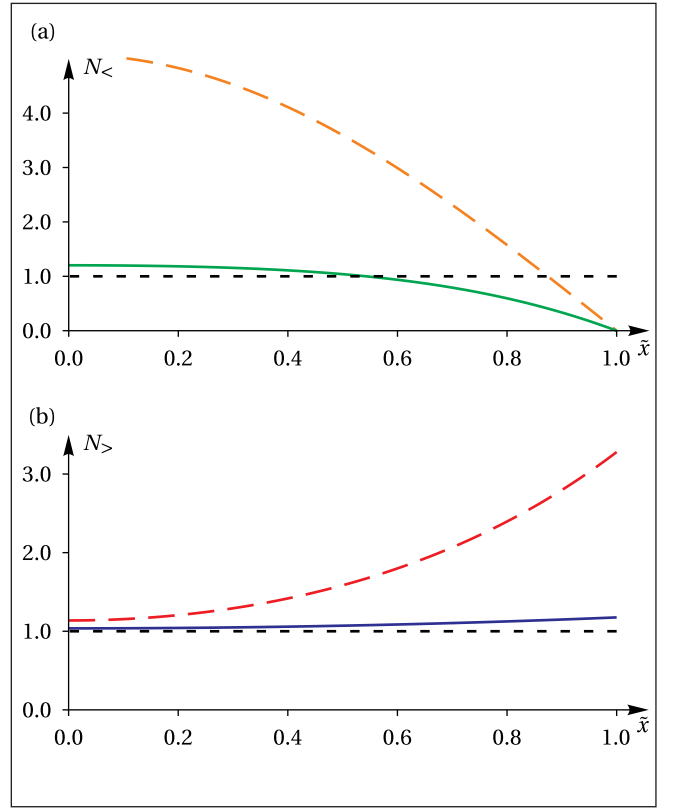}
\caption{(Color online.) Density of states as a function of the coordinate~${%
\tilde{x} = x / L}$: (a)~for ${\cos(\protect\varphi_0) < E < 1}$ with the
curves corresponding to the values ${E = 0.51}$ (long-dashed orange) and ${E
= 0.9}$ (solid green); (b)~for ${E > 1}$ with the curves corresponding to
the values ${E = 1.05}$ (long-dashed red) and ${E = 1.9}$ (solid blue). The
short-dashed black curve denotes in both cases the value ${N = 1}$. We set ${%
\cos \protect\varphi_0 = 0.5}$.}
\label{fig:Dos_on_x}
\end{figure}

\begin{figure}[tbp]
\includegraphics[width=1.0\columnwidth]{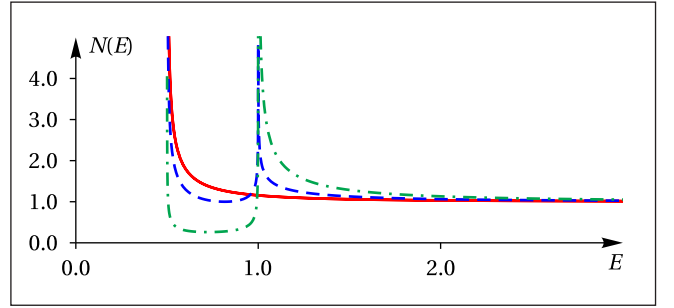}
\caption{(Color online.) Density of states as a function of the normalized
energy~${E = \protect\epsilon / \Delta}$ for different values of the
coordinate~${\tilde{x} = x / L}$, i.e., ${\tilde{x} = 0}$ (solid red), ${%
\tilde{x} = 0.5}$ (dashed blue), and ${\tilde{x} = 0.9}$ (dash-dotted
green). We set ${\cos \protect\varphi_0 = 0.5}$.}
\label{fig:DOSon_energy}
\end{figure}

We plot the DOS~$N_{<}(\epsilon, x)$ for different~${E \equiv \epsilon /
\Delta}$ in Fig.~\ref{fig:Dos_on_x}~(a). As it should be, at ${x = \pm L}$,
the DOS turns to zero. At energies~$\epsilon$ below the gap~$\tilde{\Delta}$
(${|\epsilon| \leq \tilde{\Delta}}$) in the center of the junction the DOS
is also zero. Above the gap~$\Delta$ (${|\epsilon| \geq \Delta}$) the DOS is
\begin{equation}
N_{>}(\epsilon, x) = \frac{|\epsilon|}{\sqrt{\epsilon^{2} - \tilde{\Delta}%
^{2}}} \cosh (\tilde{x} \ln M_{>}) \,,  \label{57}
\end{equation}
where
\begin{equation}
M_{>} = \frac{\Delta \sin \varphi_{0} + \sqrt{\epsilon^{2} - \Delta^{2}
\cos^{2} \varphi_{0}}}{\sqrt{\epsilon^{2} - \Delta^{2}}} \,.  \label{57'}
\end{equation}
The DOS~$N_{>}(\epsilon, x)$ for different~$E$ is shown in Fig.~\ref%
{fig:Dos_on_x}~(b).

In Fig.~\ref{fig:DOSon_energy} we plot the dependence of the DOS on the
energy ${E\equiv \epsilon /\Delta }$ for different values of~$\tilde{x}$.
In a ballistic case, the function~$N(\epsilon)$ has sharp
peaks at energies corresponding to the positions of Andreev's levels. In the
considered diffusive case these peaks are smeared out by impurity scattering
so that the dependence~$N(\epsilon)$ is a smooth curve having singularities at the edges, ${\epsilon = \Delta \cos \varphi_{0}}$ and ${\epsilon = \Delta}$.

\section{Conclusions}

We analyzed the admittance~$Y(\Omega )$ of short weak links of two types.
The first one is a one-dimensional superconducting wire with a local
suppression of the superconducting order parameter~$|\Psi |$. This system
resembles a phase-slip center or a one-dimensional
Larkin-Ovchinnikov-Fulde-Ferrell structure. We calculated~$Y$ and the
impedance ${Z=Y^{-1}}$ on the basis of non-stationary Ginzburg-Landau
equations.\cite{G-Eliash} Ac current through this wire induces a condensate
momentum~$Q_{\Omega }$ and an inhomogeneity of~$|\Psi |$ leads to branch
imbalance~$n_{\text{imb}}$ and to the appearance of another gauge-invariant
quantity, the potential~$\mu $, proportional to~$n_{\text{imb}}$.

As
we mentioned in the Introduction, the branch-imbalance, i.e., the unequal
population of the electron- and hole-like branches of the excitation
spectrum, arises in nonuniform superconductors when a conversion of the
supercurrent~$j_{S}$ into the quasiparticle current~$j_{N}$
takes place (see Refs.~\onlinecite{Tinkham_Clarke_1972,Tinkham_1972,Schmid75,ArtVolkovRev80}). The typical examples of such a conversion are the passage of the charge current through the S/N boundary,\cite{Schmid75,Artemenko1978,Ovchinnikov1978} or collective phase mode, i.e., Carlson-Goldman mode,\cite{Carlson_Goldman_1975} in uniform
superconductors.\cite{Schmid_Schoen_1975,ArtVolkov75,ArtVolkovRev80,Schoen84_a} In the latter case, nonuniform perturbations of the the currents~$j_{N}$ and~$j_{S}$ propagate with a finite wave vector~$k$ converting into each other so that the total current density ${j = j_{N} + j_{S}}$ is not perturbed, ${j = 0}$.

The electric field~$E$, which arises in the wire (see Fig.~\ref{fig:System}), is caused by both quantities,~$Q_{\Omega }$ and~$\partial_{x}\mu $ so that
neither of these quantities can be neglected (compare with a recent paper,
Ref.~\onlinecite{Dai_Lee_2017}, where only the quantity ${\mathbf{Q}\propto
\mathbf{A}}$ was taken into account). The real part of the impedance~$%
Z^{\prime }$ has a maximum at some frequency~$\Omega _{m}$ which decreases
with increasing suppression of~$|\Psi |$.

We also analyzed ac properties of short Josephson S\nobreakdash-c\nobreakdash%
-S weak links. The admittance~$Y(\Omega)$ is described by an expression that
has been obtained on the basis of microscopic equations for quasiclassical
Green's function in the Keldysh technique.\cite{AVZ79} The obtained
dependence~$Y(\Omega)$ is valid in a wide range of the frequencies~$\Omega$
and temperatures~$T$ provided the Thouless energy ${E_{\text{Th}} = D/L^{2}}$
exceeds~$\Delta$, $T$, and~$\hbar \Omega$.

At low temperatures~$T$, the absorption is absent (${Y^{\prime} = 0}$) if
the energy of photons~$\hbar \Omega$ is less than the lowest energy gap in
the center of the constriction ${\tilde{\Delta} = \Delta \cos \varphi_{0}}$.
With increasing the difference ${(\hbar \Omega - \tilde{\Delta})}$, the
absorption monotonously increases if the phase difference~$2 \varphi_{0}$ is
less than the phase difference~$2 \varphi_{\text{c}}$ corresponding to a
maximum of the Josephson current ${I_{\text{J}} = I_{\text{c}}}$. In the
interval ${2 \varphi_{\text{c}} < 2 \varphi_{0} < \pi}$, the dependence of~$%
Y^{\prime}(\Omega)$ has a maximum.

The hump in the obtained dependence~$Y^{\prime }(\Omega )$ is much broader
than the peak in absorption in a current-carrying superconductor.\cite%
{Moor_Volkov_Efetov_2017_PRL} The mechanisms causing these maxima are
different. In the first case, the maximum stems from excitation of
quasiparticles with energy range defined by~Eq.~(\ref{53'}). These
quasiparticles are bound in a potential well within the constriction. In the
second case, in Ref.~\onlinecite{Moor_Volkov_Efetov_2017_PRL}, the peak is
related to a resonance excitation of the Higgs mode by an ac field.

The anomalous enhancement of the real part of the admittance~$Y^{\prime }(\Omega )$ at low frequencies~$\Omega$
described by Eq.~(\ref{55}), is caused by interference of Cooper
pairs and quasiparticles with energies in the interval determined by Eq.~(\ref{53'}). These quasiparticles experience multiple Andreev reflections. In the ballistic case, the quasiparticles occupy Andreev's levels.\cite{Gunsenheimer_Zaikin_1994,Bratus_et_al_1995,Cuevas_et_al_1996} In the diffusive case, these levels are broadened by impurity scattering\cite{Bardas_Averin_1997,Kos_Nigg_Glazman_2013} so that the
peaks in the DOS~$N(\epsilon,x)$ corresponding to Andreev's levels
disappear, and the function~$N(\epsilon,x)$ is a smooth function
with singularities at ${\epsilon = \Delta \cos \varphi_{0}}$ and ${\epsilon = \Delta}$. In the latter case, anomalous behaviour of low energy
quasiparticles results in a singularity of dc conductance at ${V \to 0}$.\cite{Bardas_Averin_1997}

The enhancement of~$Y^{\prime }(\Omega )$ at low
frequencies results in an enhancement of the supercurrent noise because the
real part of admittance and spectral function of noise are connected by the
fluctuation-dissipation theorem. The anomalous noise in Josephson weak links
has been studied in detail in many papers.\cite{Averin_Imam_1996,Rodero_et_al_1996,Cuevas_et_al_1999,Naveh_Averin_1999,Belzig_Nazarov_2001,Bezuglyi_et_al_inbook}

Note an important circumstance. In a current-biased S\nobreakdash-c%
\nobreakdash-S JJ, the states with phase difference ${2 \varphi_{\text{c}} <
2 \varphi_{0}}$ are unstable, so that it is impossible to observe a
non-monotonous dependence of absorption in these junctions. However, in
recent experiments\cite{Dassonneville_et_al_2013,Ronzani_et_al_2016_arxiv}
it was shown that the phase difference~$2 \varphi_{0}$ in the interval
\begin{equation}
2 \varphi_{\text{c}} < 2 \varphi_{0} < \pi  \label{c1}
\end{equation}
is reachable. Thus, it would be interesting to observe a non-monotonous
dependence of absorption in such JJs by appropriate adjustment of the phase
difference~$2 \varphi_{0}$ with the help of an external magnetic field. In
these experiments, the S\nobreakdash-c\nobreakdash-S Josephson weak link was
incorporated into a superconducting loop. The phase difference is determined
by a magnetic field~$H_{0}$ through this loop,
\begin{equation}
2 \varphi_{0} = 2 \pi \Big(n + \frac{\Phi}{\Phi_{0}}\Big) \,,  \label{c2}
\end{equation}
where ${\Phi = H_{0} S + \mathcal{L} I_{\text{J}}}$ is the magnetic flux
through the loop, and $S$ respectively $\mathcal{L}$ are the area
respectively inductance of the loop.

Therefore, the absorption can be studied in the setup used in Ref.~%
\onlinecite{Ronzani_et_al_2016_arxiv} if the magnetic field contains not
only dc but also an ac component, ${H(t) = H_{0} + H_{\Omega} \cos (\Omega t)%
}$. Qualitatively, our results are applicable to the system studied in Ref.~%
\onlinecite{Ronzani_et_al_2016_arxiv} because the length of the constriction~%
$2L$ (${2L \approx 160\text{nm}}$) is comparable with the coherence length ${%
\xi_{\text{S}} \approx 100\text{nm}}$. By varying~$H_{0}$, one can change
the phase~$\varphi_{0}$ using the relation in ~Eq.~(\ref{c2}) and study the
absorption of the ac component~$H_{\Omega} \cos (\Omega t)$ as a function of
the frequency~$\Omega$.

At high temperatures (${T \gg \Delta}$), the main contribution to the
admittance~$Y_{\Omega}$ occurs due to the anomalous term. Quasiparticles
with large enough energies (${\epsilon \geq \Delta}$) yield the admittance~$%
Y^{\prime}$ approximately equal to that in the normal state ${Y_{\text{n}%
}^{\prime} = 1/R}$, whereas quasiparticles with energy range defined by~Eq.~(%
\ref{53'}) lead to an enhanced admittance~$\delta Y^{\prime}$ at low~$\Omega$
which can exceed~$Y_{\text{n}}^{\prime}$ by~$\Delta / \gamma_{\text{in}}$
times. The dependence of~$\delta Y^{\prime}$ on dc current (or phase
difference) is described by an analytical expression, Eq.~(\ref{55}).

\acknowledgments

We are grateful to Valeri V.~Pavlovskii for his kind support by numerical
calculations.

\appendix

\section{Ginzburg-Landau equation}

\label{app:GL_Eq}

Linearizing Eqs.~(\ref{1})--(\ref{4}) we obtain the set of equations for the
function $\delta E_{\Omega}$,
\begin{align}
\nu f_{0}^{2}(x) \mu_{\Omega} &= - \partial_{x} \delta E_{\Omega} \,,
\label{12} \\
\delta Q_{\Omega} f_{0}^{2}(x) + Q_{\infty} [f_{0}^{2}(x) - 1] &= - \delta
E_{\Omega} \,,  \label{12'} \\
-i \Omega \delta Q_{\Omega} &= \delta E_{\Omega} + \partial_{x} \mu_{\Omega}
\,.  \label{12''}
\end{align}

Excluding~$\delta Q_{\Omega}$ and~$\mu_{\Omega}$, we get
\begin{widetext}
\begin{equation}
\delta E_{\Omega} \nu \big[ f_{0}^{2}(x) - i \Omega \big] - \Big[ \partial_{xx}^{2} \delta E_{\Omega} - 2 \frac{\partial_{x} f_{0}(x)}{f_{0}(x)} \partial_{x} \delta E_{\Omega} \Big] = -i \Omega Q_{\infty} \nu \big[ 1 - f_{0}^{2}(x) \big] \,.  \label{13}
\end{equation}
\end{widetext}

One can exclude the first derivative~$\partial_{x} \delta E_{\Omega}$ via
the transformation ${\delta E_{\Omega} = f_{0}(x) \mathcal{E}_{\Omega}}$.
Thus, an ``effective electric field''~$\mathcal{E}_{\Omega}$ satisfies the
equation
\begin{widetext}
\begin{equation}
- \partial_{xx}^{2} \mathcal{E}_{\Omega} + \mathcal{E}_{\Omega} \Big[ \nu
\big( f_{0}^{2}(x) - i \Omega \big) + 2 \Big(\frac{\partial_{x} f_{0}(x)}{f_{0}(x)} \Big)^{2} - \frac{\partial_{xx}^{2} f_{0}(x)}{f_{0}(x)} \Big] = -i \Omega I_{\Omega} \frac{\nu \big[ 1 - f_{0}^{2}(x) \big]}{f_{0}(x)[1 - i \Omega]} \,.
\end{equation}
\end{widetext}
Using the expression Eq.~(\ref{7}) for~$f_{0}(x)$, we obtain~Eq.~(\ref{16}).

At ${x = 0}$, the function~$\mathcal{E}_{\Omega}$ is continuous, while~$%
\partial_{x} \mathcal{E}_{\Omega}$ has a jump. It can be found directly from
Eq.~(\ref{6})
\begin{align}
\big[ \partial_{x} \mathcal{E}_{\Omega} \big] &= 2 \partial_{x} \mathcal{E}%
_{\Omega}|_{0+}  \label{15a} \\
&= - \frac{\partial_{x} f_{0}(x)}{f_{0}(0)} \mathcal{E}_{\Omega}(x)|_{x=0}
\notag \\
&= - a_{0} \mathcal{E}_{\Omega}(0) \,,  \notag
\end{align}
where ${\big[ \partial_{x} \mathcal{E}_{\Omega} \big] \equiv \partial_{x} %
\big( \mathcal{E}_{\Omega}|_{x=0+} - \mathcal{E}_{\Omega}|_{x=0-} \big)}$,
see also Eq.~(\ref{BC}).

\section{Expression for the current}

\label{app:current}

In the first step, it is necessary to find the functions~$\delta \hat{a}%
^{R(A)}$. From the normalization condition~Eq.~(\ref{U3}) we get
\begin{equation}
[ \hat{a} \cdot \hat{g} + \hat{g} \cdot \hat{a} ]^{R(A)} = 0 \,.  \label{41}
\end{equation}
Linearizing this equation, we obtain for deviations caused by the ac
perturbation of the phase~$\varphi_{\Omega }$
\begin{equation}
[ \delta \hat{a} \cdot \hat{g} + \hat{g} \cdot \delta \hat{a} ]^{R(A)} = -[
\hat{a}_{0} \cdot \delta \hat{g} + \delta \hat{g} \cdot \hat{a}_{0} ]^{R(A)}
\,,  \label{42}
\end{equation}
where the matrices~$\hat{a}_{0}^{R(A)}$ are determined by Eq.~(\ref{33}). We
took into account that neither~$\hat{a}_{0}$ nor~$\delta \hat{a}$ do not
depend on the coordinate~$x$. Then, we subtract Eq.~(\ref{42}) from itself
taken at different points ${x = \pm L}$,
\begin{equation}
\big[ \delta \hat{a} \cdot \hat{G}_{\varphi}^{(-)} + \hat{G}_{\varphi}^{(-)}
\cdot \delta \hat{a} \big]^{R(A)} = - \big[ \hat{a}_{0} \cdot \delta \hat{G}%
^{(-)} + \delta \hat{G}^{(-)} \cdot \hat{a}_{0} \big]^{R(A)} \,.  \label{43}
\end{equation}

The matrix $\hat{G}_{\varphi}^{(-)}$ is defined as
\begin{equation}
\hat{G}_{\varphi}^{(-)} = \big[ \hat{G}_{\varphi}(L) - \hat{G}_{\varphi}(-L) %
\big]^{R(A)} / 2  \label{44}
\end{equation}
with ${\hat{G}_{\varphi}^{R(A)}(\pm L) = \big[G(\epsilon) \hat{\tau}_{3} +
F(\epsilon) i (\hat{\tau}_{2} \cos \varphi_{0} \pm \hat{\tau}_{1} \sin
\varphi_{0}) \big]^{R(A)}}$, and, thus, the matrix $\hat{G}{(-)}$ is given
by Eq.~(\ref{37}).

The deviation~$\delta \hat{G}^{(-)}$ due to a small phase perturbations~$%
\varphi_{\Omega}$ is found from Eq.~(\ref{30}),
\begin{align}
\delta \hat{G}^{(-)} &= \delta \big[ (1 + i \hat{\tau}_{3} \varphi_{\Omega})
\hat{G}_{\text{st}}^{(-)}(1 - i \hat{\tau}_{3} \varphi_{\Omega}) \big]
\label{45} \\
&= \varphi_{\Omega} i \big[ - \hat{\tau}_0 (G_{+} - G_{-} ) + \hat{\tau}_{1}
( F_{+} + F_{-} ) \cos \varphi_{0} \big] \,,  \notag
\end{align}
where ${G_{\pm} \equiv G_{0}(\epsilon_{\pm})}$, ${\epsilon_{\pm} = \bar{%
\epsilon} \pm \Omega / 2}$. As follows from Eq.~(\ref{26}), we need to find $%
\mathrm{Tr}\{\hat{\tau}_{3} \hat{a}\}$. Thus, multiplying Eq.~(\ref{40}) by~$%
i \hat{\tau}_{2}$ and calculating the trace, we find
\begin{widetext}
\begin{equation}
\frac{1}{2} \mathrm{Tr} \{\hat{\tau}_{3} \hat{a} \}^{R(A)} = \frac{l \varphi_{\Omega} \big\{ ( b_{+} \epsilon_{+} + b_{-} \epsilon_{-}) (G_{+} - G_{-}) - \Delta \cos^{2} \varphi_{0} (F_{+} + F_{-}) (b_{+} - b_{-}) \big\}^{R(A)} }{L (F_{+} - F_{-})^{R(A)} \sin \varphi_{0}}  \label{46} \,.
\end{equation}
\end{widetext}
The same equation holds for~$\mathrm{Tr}\{\hat{\tau}_{3} \hat{a}^{\text{an}%
}\}$ if all functions like $G_{-}^{R}$ in Eq.~(\ref{44}) for~$\hat{a}^{R}$
are replaced by~$G_{-}^{A}$.

\section{Density of states in an S-c-S~type contact}

\label{app:DOS_in_SCS}

The density of states is determined by the expression
\begin{equation}
N(\epsilon) = \frac{1}{4} \mathrm{Tr} \big\{\hat{\tau}_{3}(\hat{g}%
^{R}(\epsilon, x) - \hat{g}^{A}(\epsilon, x)) \big\} \,,  \label{D1}
\end{equation}
where~$\hat{g}^{R(A)}(\epsilon, x)$ are determined by Eq.~(\ref{27}) which
can be written in the form (dropping the indices $R(A)$)
\begin{equation}
\hat{g}(\epsilon,x) = \hat{g}(\epsilon,0) \big[ \cosh [l^{-1} x \hat{a}%
(\epsilon)] - \sinh [l^{-1} x \hat{a}(\epsilon)] \big] \,,  \label{D2}
\end{equation}
where the functions~$\cosh [l^{-1} x \hat{a}(\epsilon)]$ and~$\sinh [l^{-1}
x \hat{a}(\epsilon)]$ can be presented as follows
\begin{align}
\cosh [l^{-1} x \hat{a}(\epsilon)] &= \cosh \Big( \frac{x}{L} A \Big) \,,
\label{D3} \\
\sinh [l^{-1} x \hat{a}(\epsilon)] &= \frac{\hat{m}}{i \tilde{\zeta}%
(\epsilon)} \sinh \Big( \frac{x}{L} A \Big) \,,
\end{align}
where we defined ${A = \mathrm{arsinh} \big[ \zeta^{-1}(\epsilon) \Delta
\sin \varphi_{0} \big] = \ln M}$ with ${M = \zeta^{-1}(\epsilon) \big[ %
\Delta \sin \varphi_{0} + \tilde{\zeta}(\epsilon) \big]}$, and ${\hat{m} =
\tilde{\Delta} \hat{\tau}_{3} + \epsilon i \hat{\tau}_{2}}$. We used the
expression for the matrix $\hat{a}$ from Ref.~\onlinecite{AVZ79},
\begin{equation}
\hat{a} = - \frac{il}{L \tilde{\zeta}(\epsilon)} \hat{m} A \,.
\end{equation}
The matrix~$\hat{g}(\epsilon, 0)$ is found from Eq.~(\ref{28}),
\begin{equation}
\hat{g}(\epsilon, 0) = \frac{\hat{G}^{(+)}}{\cosh A} \,.  \label{D4}
\end{equation}
Therefore, Eq.~(\ref{D2}) can be written in the form
\begin{equation}
\hat{g}(\epsilon,x) = \frac{\hat{G}_{+}}{\cosh A} \Big[ \cosh( \tilde{x} A )
+ \frac{\hat{m}}{i \tilde{\zeta}(\epsilon)} \sinh ( \tilde{x} A ) \Big] \,.
\label{D5}
\end{equation}
Using Eq.~(\ref{D1}) we obtain for the density of states
\begin{equation}
N(\epsilon, x) = \frac{1}{2} \Bigg\{ \Big[ g(\epsilon) \frac{\cosh(\tilde{x}
\ln M)}{\cosh(\ln M)} \Big]^{R} - \Big[ g(\epsilon) \frac{\cosh (\tilde{x}
\ln M)}{\cosh (\ln M)} \Big]^{A} \Bigg\} \,.  \label{D6}
\end{equation}
One can easily show that ${\cosh (\ln M) = \tilde{\zeta}(\epsilon)
\zeta^{-1}(\epsilon)}$. Therefore, the density of states~$N(\epsilon, x)$
for ${|\epsilon| \geq \Delta}$ when ${\tilde{\zeta}^{R}(\epsilon) = - \tilde{%
\zeta}^{A}(\epsilon)}$ can be written as follows
\begin{equation}
N(\epsilon, x) = \frac{|\epsilon|}{2 \tilde{\zeta}(\epsilon)} \big\{ \big[%
\cosh (\tilde{x} \ln M) \big]^{R} + \big[ \cosh (\tilde{x} \ln M) \big]^{A} %
\big\} \,.  \label{D7}
\end{equation}

Consider two cases:

\begin{enumerate}
\item ${|\epsilon| \geq \Delta}$. In this case, ${M^{R}= \big[ M^{A} \big]%
^{-1} \equiv M_{>}}$, where ${M_{>} = \zeta_{>}^{-1}(\epsilon) \big[ \Delta
\sin \varphi_{0} + \tilde{\zeta}_{>}(\epsilon) \big]}$ with ${\tilde{\zeta}%
_{>}(\epsilon) = \sqrt{\epsilon^{2} - \tilde{\Delta}^{2}}}$, and ${%
\zeta_{>}(\epsilon) = \sqrt{\epsilon^{2} - \Delta^{2}}}$. We obtain
\begin{align}
N(\epsilon, x) &= \frac{|\epsilon|}{\tilde{\zeta}(\epsilon)} \cosh \big(
\tilde{x} \ln M_{>} \big )  \label{D8} \\
&= \frac{|\epsilon|}{2 \tilde{\zeta}(\epsilon)} \big[ M_{>}^{\tilde{x}} +
M_{>}^{-\tilde{x}} \big] \,.  \notag
\end{align}

\item ${\tilde{\Delta} \leq |\epsilon| \leq \Delta}$. In this case, ${M^{R}
= -i\zeta_{<}^{-1}(\epsilon) \big[ \Delta \sin \varphi_{0} + \tilde{\zeta}%
_{>}(\epsilon) \big] \equiv -i M_{<}}$ and ${M^{A} = -i
\zeta_{<}^{-1}(\epsilon) [\Delta \sin \varphi_{0} - \tilde{\zeta}%
_{>}(\epsilon)]}$, and one can write the sum in Eq.~(\ref{D7}) in the form
\begin{widetext}
\begin{align}
\big[\cosh (\tilde{x} \ln M) \big]^{R} + \big[ \cosh (\tilde{x} \ln M) \big]^{A} &= 2 \cosh \Big[ \frac{\tilde{x}}{2} \big( M^{R} + M^{A} \big) \Big] \cosh \Big[ \frac{\tilde{x}}{2} \big( M^{R} - M^{A} \big) \Big] \label{D9} \\
&= 2 \cos \Big(\frac{\pi \tilde{x}}{2}\Big) \cosh \big[ \tilde{x} \ln M_{<} \big] \,.
\end{align}
\end{widetext}
\end{enumerate}

Combining Eqs.~(\ref{D7})--(\ref{D9}) we obtain Eq.~(\ref{56a}).


%

\end{document}